\documentclass[]{jfm}

\usepackage{setspace}
\usepackage{amsmath}
\usepackage{listings}
\usepackage{xcolor}

\usepackage{subcaption}

\usepackage{graphicx}
\usepackage{booktabs}
\usepackage{newtxtext}
\usepackage{newtxmath}
\usepackage{natbib}
\usepackage{hyperref}
\hypersetup{
    colorlinks = true,
    urlcolor   = blue,
    citecolor  = black,
}

\newcommand{\RomanNumeralCaps}[1]
\linenumbers

\usepackage{multirow}

\newcommand{\celsius}{\,^\circ\text{C}}
\newcommand{\rhoMD}{\rho_\text{MD}}
\newcommand{\TMD}{T_\text{MD}}
\newcommand{\erfc}[1]{\text{erfc}\left({#1}\right)}
\newcommand{\erf}[1]{\text{erf}\left({#1}\right)}
\newcommand{\proot}[2]{\sqrt[\leftroot{-1}\uproot{10}#2]{#1}}

\newcommand{\Ra}{\text{Ra}}
\renewcommand{\Pr}{\text{Pr}}

\newcommand{\mat}[1]{\mathsfbi{#1}}

\renewcommand{\vec}[1]{\boldsymbol{#1}}

\graphicspath{{figures/}}

\title{Onset of Radiatively Driven Convection in Ice-Covered Freshwater Lakes}

\author{Benjamin J. Smith\aff{1,2}
  \corresp{\email{bjsmith28@wisc.edu}},
  Bingfei Yan\aff{1}
  \corresp{\email{byan44@wisc.edu}},
  Till J. W. Wagner\aff{2}, 
  Nimish Pujara\aff{3,4},
 \and Jennifer A. Franck\aff{1}}

\affiliation{\aff{1}Department of Mechanical Engineering, University of Wisconsin-Madison, Madison, WI, USA
\aff{2}Department of Atmospheric \& Oceanic Sciences, University of Wisconsin-Madison, Madison, WI, USA
\aff{3}Department of Civil \& Environmental Engineering, University of Wisconsin-Madison, Madison, WI, USA
\aff{4}Department of Civil Engineering, University of British Columbia, Vancouver, BC, Canada}

\begin{document}

%%%%%%%%%%%%%%%%%%%%%%%%%%%%%%%%%%%%%%%%%%%%%%%%%%%%%%%%%%%%%%%%

\maketitle

% All papers should feature a single-paragraph abstract of no more than 250 words which must not spill onto the second page of the manuscript
\begin{abstract}
    This work investigates the formation and growth of Rayleigh-Taylor instabilities (RTIs) during the onset of radiatively driven convection (RDC) in seasonally ice-covered freshwater lakes. Two-dimensional direct numerical simulations (DNS) and linear analysis allow for the investigation of how environmental factors, including thermal stratification, water clarity, and solar forcing influence RTI formation. It is found that for strongly stratified lakes, there is a period of dormancy before the water column forms a gravitationally unstable layer that allows RTI growth. Conversely, for weakly stratified lakes the water column can become unstable almost immediately, allowing for rapid RTI growth. A linear analysis is performed to investigate the early RTI growth dynamics, and its results compare well with DNS. Notably, the wavelengths and depths of RTIs, and their associated time-dependent growth rates, are all captured by the linearised dynamics. Predictions of RDC onset obtained using the linear model show good agreement with DNS results. When the analysis is applied to field measurements of ice-covered lakes, it is found that weakly stratified lakes may take a few hours before RDC onset whereas strongly stratified lakes may not see RDC onset after a full day of clear-sky spring conditions.
\end{abstract}

% Authors should not enter keywords on the manuscript
\begin{keywords}

\end{keywords}

%{\bf MSC Codes }  {\it(Optional)} Please enter your MSC Codes here

%%%%%%%%%%%%%%%%%%%%%%%%%%%%%%%%%%%%%%%%%%%%%%%%%%%%%%%%%%%%%%%%
%%%%%%%%%%%%%%%%%%%%%%%%%%%%%%%%%%%%%%%%%%%%%%%%%%%%%%%%%%%%%%%%
\section{Introduction}
\label{sec:intro}
%%%%%%%%%%%%%%%%%%%%%%%%%%%%%%%%%%%%%%%%%%%%%%%%%%%%%%%%%%%%%%%%

This study mathematically models the early development of Rayleigh-Taylor instabilities (RTIs) that are necessary to drive radiatively driven convection (RDC) underneath seasonally ice-covered freshwater lakes. A linearised framework provides predictions for how environmental conditions (e.g. solar radiation) impact the timing, structure, and growth rates of these instabilities. 

\begin{figure}
    \centering
    \includegraphics{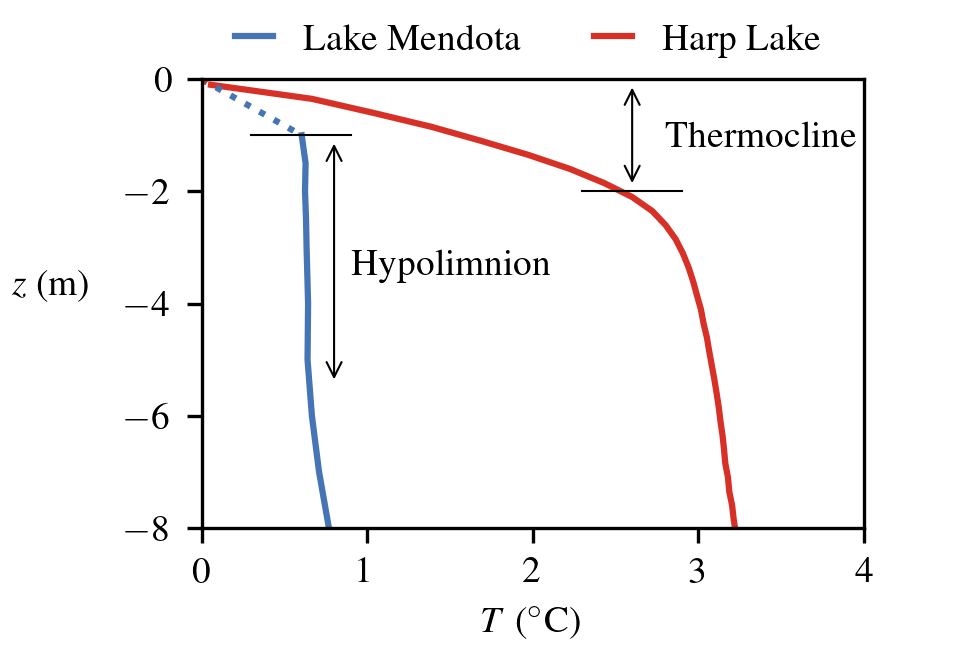}
    \caption{\label{fig:lake-profiles} Measured temperature profiles averaged over an early spring day (1st March) for ice-covered Lake Mendota, WI, USA in 2019 \citep{lottig2022} and ice-covered Harp Lake, ON, Canada in 2014 \citep{wells2022}. Note measurements are not available above the first meter for Lake Mendota and are generated by linear extrapolation (dotted lines) assuming the temperature is $0\celsius$ at the ice-water interface ($z=0$ m).}
\end{figure}

Many freshwater lakes located within the northern hemisphere frequently freeze over during the winter period, causing an inverse thermal stratification \citep{Kirillin2012,yang2021}. Due to properties of cold freshwater, where increases in temperatures between $0\celsius$ and $4\celsius$ increase water density, the water column is in a state of gravitational stability suppressing water motion \citep{MATTHEWS1987,Bouffard2019a}. Two representative profiles of water temperatures in ice-covered lakes are shown in figure~\ref{fig:lake-profiles}. However, this stable thermal stratification can be disturbed by buoyancy-driven flows such as RDC. This tends to occur in the early spring once light penetrates through the snow and ice to heat near-surface waters and overcome density gradients to promote the growth of RTIs \citep{allum2023thesis, Smith2025}. Eventually, a turbulent `convective mixed layer' develops near the ice-water interface that transports heat and particulates in the upper water column \citep{Mironov2002,Bouffard2019a,Jansen2021}. This can facilitate early algal blooms \citep{Kelley1997}, which have been found to become more frequent even in cold waters \citep{reinl2023}. Further, climate change is driving shifts in environmental conditions, for example the duration and composition of the ice cover \citep{Weyhenmeyer2022,culpepper2024,salamattalab2026}, which will impact RDC and subsequently lake ecology. 

Although RDC has been observed in freshwater lakes for both ice-free \citep[e.g.][]{austin2022} and ice-covered conditions \citep[e.g.][]{Farmer1975,Mironov2002,Jonas2003,Bouffard2019b,kirillin2026}, in situ observations during ice-covered periods are made difficult by adverse winter conditions \citep{Ozersky2021}. Small scale experiments have been conducted but have focused on the later mixing dynamics in warm water rather than the initial RTI growth phase in colder climates \citep{Lepot2018,Bouillaut2019,hadjerci2025}. As a result, recent progress in our understanding has been driven primarily by theoretical and numerical studies across a range of scales, from lake-wide systems \citep[e.g.][]{Ulloa2019} to smaller process-specific studies \citep[e.g.][]{Allum2022}. Recent simulations by \citet{Smith2025} focused on RDC onset within the thermocline: a layer of increasing temperature found between the ice cover and a deeper layer of warmer water of approximately uniform temperature \citep{Leppaeranta2015} referred to as the hypolimnion. \citet{Smith2025} found three regimes: (i) a period of time ($t<t_0$) where the water column is quiescent and solar forcing drives the formation of a gravitationally unstable layer; (ii) a period ($t_0<t<t_1$) where the water column is no longer stably stratified and small perturbations can begin to grow to form RTIs; and (iii) a period ($t>t_1$) where RTIs have matured at time $t_1$ to initiate RDC. Beyond $t_1$ the convective mixed layer grows in time (under continued forcing). However, \citet{Smith2025} did not provide any theoretical understanding of RTIs, such as how to predict $t_1$ or consider how they are modified by interactions with the deeper hypolimnion and other environmental factors such as the water clarity. 

Pre-winter conditions are known to impact late-winter environmental conditions \citep{Kirillin2012} 
 including water clarity \citep{gushulak2025,zhang2025}, stratification \citep{yang2021}, and ice formation \citep{culpepper2024}. For instance, \citet{Allum2022} and \citet{Allum2024} found that horizontal variations in light transmission, for example due to partial snow cover, lead to horizontal currents in addition to vertical transport. In particular, \citet{Allum2022} varied both the initial water temperature and water clarity, finding that increases in either lead to a delay in RTI formation ($t_1$). However, the authors focused on the dynamics at some depth below the surface in an unstratified water column, neglecting the roles of both the thermocline and the diffusive thermal boundary layer at the ice-water interface. \citet{allum2023thesis} included the effect of stratification, and \citet{Smith2025} combined both effects to study the impact of ice composition on RDC development. They showed how reduced light transmittance with increased ice opacity delays the formation of the unstable layer ($t_0$) and subsequently RDC onset. The effect of sediment heating was incorporated by \citet{estay2026} who then classified distinct convective regimes and their corresponding heat flux into the ice. However, no study has yet considered the influence of both the thermocline and hypolimnion, or interactions between the two. 

In this work, we provide a general framework that enables a unified study of ice-covered lakes with a generic thermal structure. First, two-dimensional, high resolution direct numerical simulations (DNS) are performed to observe and analyse how varying the thermal structure, solar-forcing, and relative water clarity affects RTI growth and RDC onset. Next, a linear analysis is applied to understand and predict the RTI growth phase including the wavelengths and depths of RTIs, their associated time-dependent growth rates, and time of onset ($t_1$). Although nonlinearity dominates after RDC onset ($t>t_1$), the dynamics prior to $t_1$ may be approximated by linear analysis \citep{chang2023,yang2026}. Prior work on RDC has implemented linear analysis for both ice-free \citep{chang2023,yang2026} and ice-covered \citep{huang2025} lakes. However, in the latter case, the authors considered a steady analysis with an initial mostly unstable water column, whereas we consider an initially stable stratification that becomes unstable over time. 

To study the impact of more complex thermal structures, we incorporate the classification of seasonally ice-covered lakes of \citet{yang2021}. They distinguish between (i) cryomictic lakes (e.g. Lake Mendota, figure~\ref{fig:lake-profiles}) characterised by a shallow thermocline ($\lesssim1$ m) and a cold hypolimnion ($<2\celsius$) formed by strong winds prior to ice-on; and (ii) cryostratified lakes (e.g. Harp Lake, figure~\ref{fig:lake-profiles}) characterised by a deeper thermocline ($>1$ m) and a warmer hypolimnion ($>2\celsius$) formed by little-to-no wind prior to ice-on. Although both regimes share similarities in their general profile, the differences in thermal structure will affect the dynamics during RDC onset. Specifically, the differences in thermocline depth and temperature will impact the location and timescales ($t_0$, $t_1$) associated with RTI growth. For example, cryomictic lakes are expected to have significant heating within the unstratified hypolimnion, which leads to a fast onset of gravitational instability ($t_0\approx0$) and quick RTI growth, similar to studies that consider unstratified domains \citep[e.g.][]{Allum2022}. Conversely, for cryostratified lakes, heating occurs within the stratified thermocline, and so the dynamics are similar to other stratified studies \citep[e.g.][]{Smith2025} in which there is a non-negligible period of time for an unstable layer to develop ($t_0$) delaying RTI growth.  

Our primary goals are to determine how the cryomictic and cryostratified thermal structures and changes in solar-forcing affect the RTI growth dynamics and resultant structure and onset of RDC. The model equations and scalings appropriate for analysing RTIs are outlined in section~\ref{sec:modelsec}. The numerical simulations are described in section~\ref{sec:dns} with corresponding results in section~\ref{sec:dns-results}. The linear analysis is detailed in section~\ref{sec:linear}, enabling the predictions of RDC onset ($t_1$), which we then apply to the case studies of Lake Mendota and Harp Lake. Finally, the conclusions of this work are presented in section~\ref{sec:conc}. 

%%%%%%%%%%%%%%%%%%%%%%%%%%%%%%%%%%%%%%%%%%%%%%%%%%%%%%%%%%%%%%%%
%%%%%%%%%%%%%%%%%%%%%%%%%%%%%%%%%%%%%%%%%%%%%%%%%%%%%%%%%%%%%%%%
\section{Modelling Radiatively Driven Convection Under Ice}
\label{sec:modelsec}
%%%%%%%%%%%%%%%%%%%%%%%%%%%%%%%%%%%%%%%%%%%%%%%%%%%%%%%%%%%%%%%%

\subsection{Equations of Motion}\label{sec:model}

\begin{figure}
    \centering
    \includegraphics[width=0.9\linewidth]{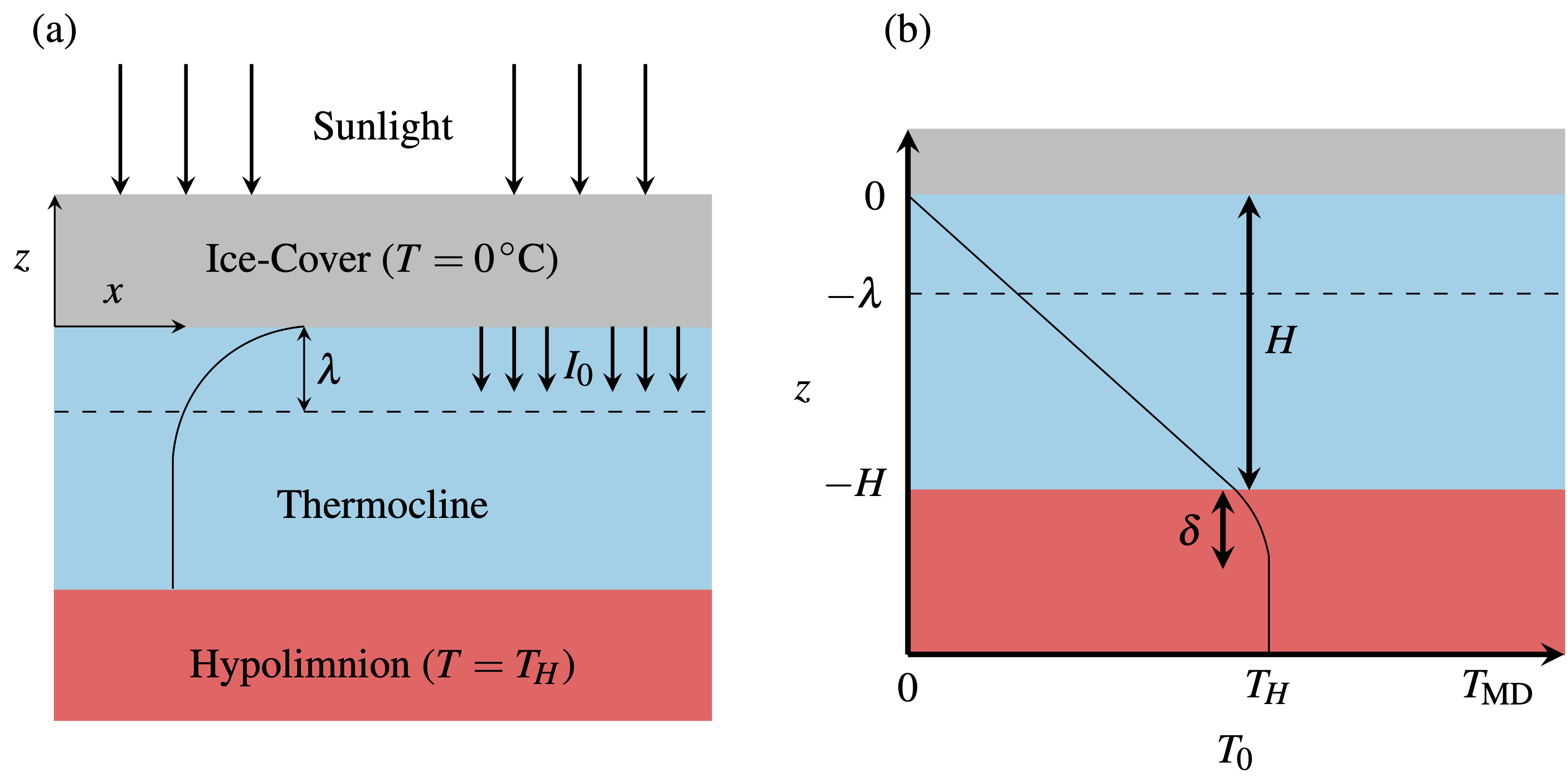}
    \caption{\label{fig:cartoon} Schematic of the under-ice water column. (a) A thermocline of depth $H$ separates the ice cover at $T = 0\celsius$ from the warmer hypolimnion at temperature $T_H\leq\TMD$. Sunlight propagates through, and is attenuated by, the ice cover. The under-ice light intensity $I_0$ then decays exponentially according to the attenuation depth $\lambda$ of the water. (b) The initial temperature $T_0(z)$ increases linearly with depth within the thermocline before transitioning over a distance $\delta$ to the isothermal hypolimnion.}
\end{figure}

To simplify the problem, an idealised model is constructed (figure~\ref{fig:cartoon}) where we treat the ice cover as an isothermal ceiling at fixed temperature $0 \celsius$ and do not consider ice melt or growth. Incoming solar radiation is assumed monochromatic, short wave and strong enough to penetrate through the ice and heat the water column. The composition of the ice cover can control the amount of transmitted light, or equivalently thermal forcing, applied to the water column \citep{Smith2025}. Here we consider only the light intensity at the ice-water interface ($I_0$), which in reality is a function of solar irradiance above the ice, the snow cover and ice composition. 

Below the ice cover, the Beer-Lambert law ($I(z) = I_0e^{z/\lambda}$; figure~\ref{fig:cartoon}a) is assumed to govern the light availability in the water column. In this model, light decays exponentially with depth (note $z<0$) according to the attenuation length of the water, $\lambda$, which characterises the water clarity. For example, highly turbid water exhibits $\lambda\sim 0.1\,\text{m}$ whereas mostly clear water has $\lambda\sim 1\,\text{m}$ \citep{Grace2022}. 

As light is absorbed, it is transformed into thermal energy and heats the water. For freshwater bodies, the water density $\rho$ is primarily a function of temperature ($T$) and for the temperature range considered here it is well approximated by the quadratic form \citep{Grace2022}
\begin{equation}\label{eq:density}
    \rho(T) = \rhoMD[1 - C(T-\TMD)^2],
\end{equation}
where $\TMD\approx3.98\celsius$ is the temperature of maximum density ($\rhoMD\approx999.975$ kg m$^{-3}$) and $C=8.3\times10^{-6}\,\celsius^{-2}$ is a fitting constant that acts as a thermal expansion coefficient. Thus, for $T<\TMD$, increasing $T$ increases the density, which provides the conditions necessary for RDC. 

The water motion is modelled by the incompressible Navier-Stokes equations under the Boussinesq approximation with radiative heating,
\begin{subequations}\label{eq:NSD}
    \begin{gather}
        \bnabla\bcdot\vec{u} = 0, \\[5pt]
        \frac{\partial\vec{u}}{\partial t} + \vec{u}\bcdot\bnabla\vec{u} = -\frac{1}{\rhoMD}\bnabla p + \nu\nabla^2\vec{u} -\frac{\rho(T)g}{\rhoMD}\vec{\hat{e}}_z, \\[5pt] 
        \label{eq:heatD}
        \frac{\partial T}{\partial t} + \vec{u}\bcdot\bnabla T = \kappa\nabla^2T + \frac{Q_0}{\lambda}e^\frac{z}{\lambda},
    \end{gather}
\end{subequations}
where $\vec{u}$, $p$, $\nu$ and $\kappa$ are the respective  velocity, pressure, viscosity and thermal diffusivity of the water, and $g$ is the acceleration due to gravity. The last term of equation~\eqref{eq:heatD} is a depth-dependent source term and represents the radiative heating according to the Beer-Lambert law. The parameter $Q_0\equiv I_0/(\rhoMD c_p)$ is the magnitude of heating at the ice-water interface where $c_p$ is the heat capacity of cold water. 

The water column is assumed initially quiescent and the thermal structure is decomposed into a linearly increasing thermocline that smoothly transitions to an isothermal hypolimnion at temperature $T_H$ located at a depth $H$ below the surface (figure~\ref{fig:cartoon}b). This idealised initial state is
\begin{equation}\label{eq:DICs}
    \vec{u}(t=0) = 0, \quad T(t=0) = \frac{T_H}{2H}\int_z^0 1 + \tanh{\left(\frac{z'+H}{\delta}\right)}\; dz',
\end{equation}
where $\delta$ controls the radius of curvature at the thermocline--hypolimnion transition. 

The boundary conditions are assumed periodic in the horizontal direction ($x$). At large depths the water is undisturbed from its initial state \eqref{eq:DICs} so that the boundary conditions in the vertical direction ($z$) are
\begin{subequations}\label{eq:DBCs}
    \begin{gather}
        \vec{u}(z=0) = \vec{u}(z\rightarrow-\infty) = 0, \\[5pt]
        T(z=0) = T(z\rightarrow-\infty) - T_H = 0.
    \end{gather}
\end{subequations}

%%%%%%%%%%%%%%%%%%%%%%%%%%%%%%%%%%%%%%%%%%%%%%%%%%%%%%%%%%%%%%%%
%%%%%%%%%%%%%%%%%%%%%%%%%%%%%%%%%%%%%%%%%%%%%%%%%%%%%%%%%%%%%%%%
\subsection{Scaling the Equations of Motion}
\label{sec:scaling}
%%%%%%%%%%%%%%%%%%%%%%%%%%%%%%%%%%%%%%%%%%%%%%%%%%%%%%%%%%%%%%%%

To scale the equations of motion \eqref{eq:NSD}, it is assumed that the pressure ($p$) is in balance with the hydrostatic pressure of maximum density $P_\text{MD}=-\rhoMD gz$ so that $p=P+P_\text{MD}$. All physical variables are then scaled by characteristic values as follows:
\begin{equation}\label{eq:scalings}
    T^* = T_\lambda T + \TMD, \quad \vec{u}^* = W\vec{u}, \quad P^* = \rhoMD W^2P, \quad  t^* = \tau t, \quad \vec{x}^* = \lambda \vec{x},
\end{equation}
where $\vec{x}=(x,z)$ and, hereafter, asterisks refer to dimensional quantities. 

The characteristic temperature scale, $T_\lambda$, is taken from the initial background temperature \eqref{eq:DICs} at the depth where initial perturbations are expected to grow. This depends on the ratio $\lambda/H$ which determines the location or spatial concentration of heating in the water column. If $\lambda/H \gtrsim 1$, radiation interacts with the isothermal hypolimnion and RTI development is expected at temperatures close to the far-field temperature $T_H$ \citep{chang2023}. Conversely, if $\lambda/H \ll 1$, radiation is absorbed predominantly within the linear thermocline, suggesting RTI development close to the surface where temperatures scale as $(\lambda/H)T_H$ \citep{Smith2025}. Consequently, we write
\begin{equation}
    T_\lambda \equiv \left(\frac{\lambda}{H}\right)^\sigma T_H \quad \mathrm{where} \quad \sigma = \begin{cases} 0, & \lambda/H\geq1, \\ 1, & \lambda/H<1. \end{cases}
\end{equation}
Following \citet{chang2023}, the characteristic velocity scale during the initial linear growth phase is found to be $W\equiv(2\lambda CgQ_0\Delta T)^{1/3}$ with time scale $\tau\equiv\lambda/W$ (see Appendix~\ref{app:LA}) where $\Delta T\equiv\TMD-T_\lambda$ defines the difference between $\TMD$ and the background temperature $T_\lambda$.  

Inserting \eqref{eq:scalings} into \eqref{eq:NSD} yields the non-dimensional equations
\begin{subequations}\label{eq:NS}
    \begin{gather}
        \bnabla\bcdot\vec{u} = 0, \\[5pt]
        \frac{\partial\vec{u}}{\partial t} + \vec{u}\bcdot\bnabla\vec{u} = -\bnabla P + \sqrt{\frac{\Pr}{\Ra}}\nabla^2\vec{u} + \frac{\beta}{2}\gamma\sqrt{\Ra\Pr}\left(\frac{\lambda}{H}\right)^{\sigma-1}T^2\vec{\hat{e}}_z, \label{eq:NSmom} \\[5pt] 
        \label{eq:heatND}
        \frac{\partial T}{\partial t} + \vec{u}\bcdot\bnabla T = \frac{1}{\sqrt{\Ra\Pr}}\nabla^2T + \frac{1}{\gamma\sqrt{\Ra\Pr}}\left(\frac{\lambda}{H}\right)^{1-\sigma}e^z,
    \end{gather}
\end{subequations} 
subject to the scaled boundary and initial conditions 
\begin{subequations}\label{eq:BCs}
    \begin{gather}
        \vec{u}(z=0) = \vec{u}(z\rightarrow-\infty) = 0, \\[5pt]
        T(z=0) + \TMD/T_\lambda = T(z\rightarrow-\infty) - (T_H-\TMD)/T_\lambda = 0, \label{eq:InfBC} \\[5pt]
        \vec{u}(t=0) = 0, \quad T(t=0) = \frac{1}{2}\left(\frac{\lambda}{H}\right)^{1-\sigma}\int_z^0 1 + \tanh{\left(\frac{c_1z'+1}{c_2}\right)}\; dz' - \frac{\TMD}{T_\lambda}, \label{eq:IC}
    \end{gather}
\end{subequations}
where $c_1=\lambda/H$ and $c_2=\delta/H$. 

Here, the Rayleigh number measures the strength of buoyancy to diffusion,
\begin{equation}\label{eq:Ra}
    \Ra \equiv \frac{\lambda^2W^2}{\nu\kappa} = \frac{(2CgQ_0\Delta T)^{2/3}\lambda^{8/3}}{\nu\kappa}.
\end{equation}
Previously reported Rayleigh numbers range from $O(10^3)$ for the development of RTIs to $>10^6$ for fully developed turbulent mixing \citep{Bouffard2019a}. For typical late-winter and early-spring conditions, we estimate $\Ra \sim O(10^5$--$10^7)$, which sets the parameter range examined here \citep[similar values have been considered previously, e.g.][]{Allum2022}.  

The Prandtl number is $\Pr\equiv\nu/\kappa\approx10$ for cold water and the forcing parameter (or equivalently the inverse of the radiative-conductive number) is defined as 
\begin{equation}
    \gamma \equiv \frac{T_H\kappa}{H Q_0},
\end{equation}
which measures the ratio of thermal diffusion to external solar forcing \citep{Smith2025,estay2026}. We estimate $\gamma \sim O(10^{-1}$--$10^{-3})$, where smaller values represent stronger relative forcing. Lastly, $\beta\equiv T_\lambda/\Delta T$ is the background temperature relative to its difference from $\TMD$. 

\begin{figure}
    \centering
    \includegraphics{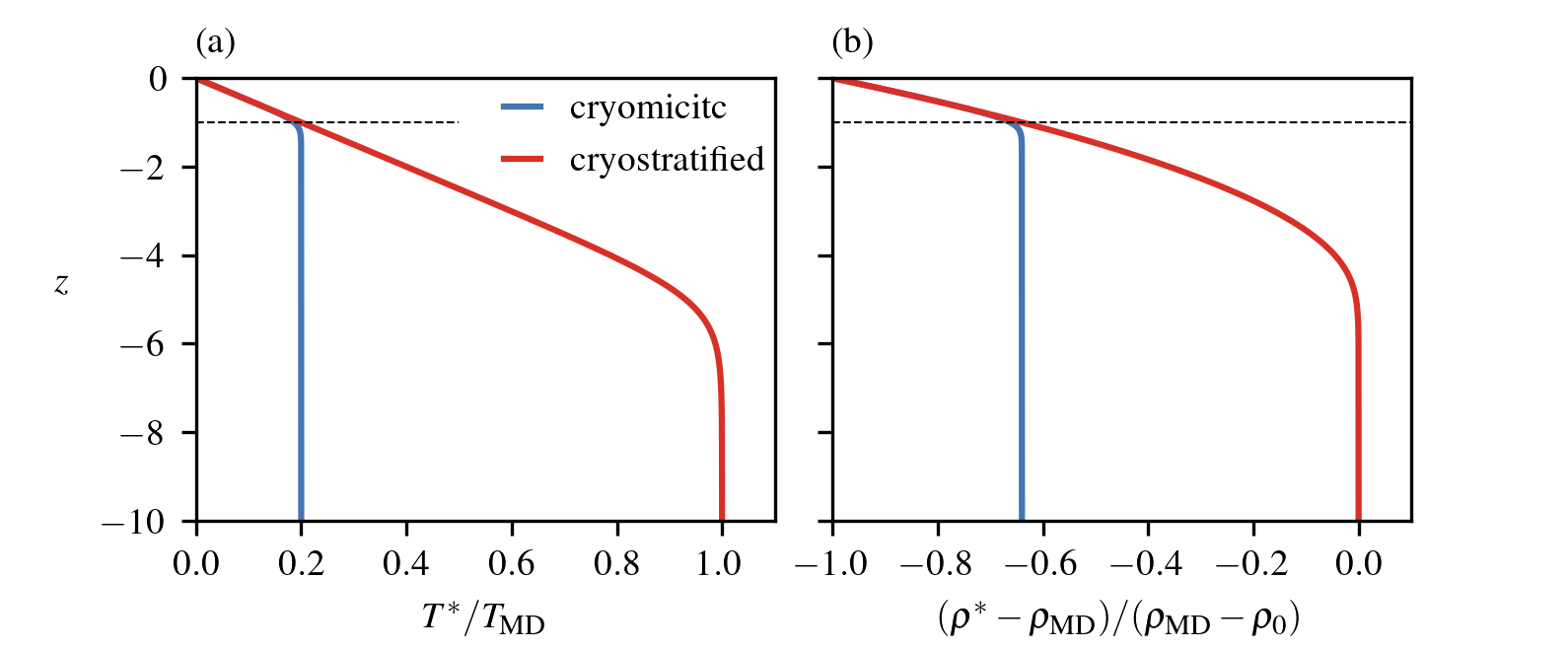}
    \caption{\label{fig:ICs} Initial profiles with $\delta/H=0.2$ for (a) water temperature and (b) water density for a cryomictic case ($\lambda/H=1$, $T_H/\TMD=0.2$) and a cryostratified case ($\lambda/H=0.2$, $T_H/\TMD=1$). The dashed line refers to the depth $z=-\lambda$ where there is significant light absorption and heating. Note that $\rho_0\equiv\rho(0\celsius)=\rhoMD(1-C\TMD^2)$.}
\end{figure}

Within this framework, cryomictic and cryostratified lakes are characterised by $\lambda/H$ (location of heating) and by $T_H/\TMD$ (hypolimnion temperature). Cryomictic lakes (e.g. Lake Mendota) are characterised by a shallow and cold hypolimnion, thus we choose $\lambda/H=1$ and $T_H/\TMD=0.2$ as representative parameters (figure~\ref{fig:ICs}a, blue line). Whereas cryostratified lakes (e.g. Harp Lake) are characterised by a deeper and much warmer hypolimnion, and so we choose $\lambda/H=0.2$ and $T_H/\TMD=1$ instead (figure~\ref{fig:ICs}a, red line). Note that as $T_H/\TMD\rightarrow1$, the density $\rho$ behaves increasingly nonlinearly with depth (figure~\ref{fig:ICs}b), due to the quadratic equation of state \eqref{eq:density}. For both the cryomictic and cryostratified regimes, $\Delta T\equiv0.8\TMD$, and $\lambda/H \leq 1$, so that $(\lambda/H)^{1-\sigma} \equiv 1$, meaning $T_\lambda/\TMD=0.2$ and $\beta=0.25$. 

Lastly, the parameter $\delta/H$ refers to the relative radius of curvature of the temperature profile as it smoothly transitions from the thermocline to the hypolimnion (figure \ref{fig:cartoon}). We expect this in most cases to be reasonably small such that $\delta/H\lesssim0.4$. 

Table~\ref{tab:cases} lists the non-dimensional parameters varied in the numerical simulations of section~\ref{sec:dns}. These represent how changes in relative water clarity, stratification, and solar forcing interact to encourage or suppress the formation of the unstable layer and subsequent RTIs. 

%%%%%%%%%%%%%%%%%%%%%%%%%%%%%%%%%%%%%%%%%%%%%%%%%%%%%%%%%%%%%%%%
%%%%%%%%%%%%%%%%%%%%%%%%%%%%%%%%%%%%%%%%%%%%%%%%%%%%%%%%%%%%%%%%
\section{Numerical Simulations}
\label{sec:dns}
%%%%%%%%%%%%%%%%%%%%%%%%%%%%%%%%%%%%%%%%%%%%%%%%%%%%%%%%%%%%%%%%

To observe the development of RTIs, the system \eqref{eq:NS} is directly solved numerically in two dimensions ($x,z$). Although three-dimensional effects (e.g. vortex stretching) become important during the nonlinear convective mixing regime \citep{allum2025}, we do not consider three-dimensions to reduce computational cost and because instabilities are expected to initially develop in two dimensions (Supplementary Material S1). To solve \eqref{eq:NS} we follow the numerical approach of \cite{Smith2025}, using the spectral solver Dedalus which implements sparse, high order spectral methods to fully resolve the flow \citep{Burns2020}. 

\begin{table}
	\centering
	\setlength{\tabcolsep}{12pt} % Adjusts horizontal column padding
	\begin{tabular}{c c c c}
		Configuration & $\text{Ra}$ & $\gamma$ & $\delta/H$ \\
		\midrule
		Baseline & $10^6$ & $10^{-2}$ & $0.2$ \\
		High $\gamma$ & $10^6$ & $\mathbf{10^{-1}}$ & $0.2$ \\
		Low $\gamma$ & $10^6$ & $\mathbf{10^{-3}}$ & $0.2$ \\
		Low $\Ra$ & $\mathbf{10^5}$ & $10^{-2}$ & $0.2$ \\
		High $\Ra$ & $\mathbf{10^7}$ & $10^{-2}$ & $0.2$ \\
		Low $\delta/H$ & $10^6$ & $10^{-2}$ & $\mathbf{0.1}$ \\
		High $\delta/H$ & $10^6$ & $10^{-2}$ & $\mathbf{0.4}$ \\
	\end{tabular}
	\caption{\label{tab:cases} Parameter space for the numerical simulations in section~\ref{sec:dns}. All parameters are evaluated across both cryomictic ($\lambda/H = 1, T_H/T_{\text{MD}} = 0.2$) and cryostratified ($\lambda/H = 0.2, T_H/T_{\text{MD}} = 1$) regimes. Bold text represents the varied parameter for each configuration.}
\end{table}

The far field bottom boundary of the domain is set at $10\lambda$, with a domain width of $20\lambda$. The domain is discretised using $M=2048$ Chebyshev polynomials in the $z$ direction and $2M$ Fourier modes in the $x$ direction which are both de-aliased by $3/2$. Fourier series in $x$ are implemented since they naturally capture periodic boundary conditions whilst Chebyshev nodes cluster at boundaries which helps resolve the boundary layer at the ice-water interface. The domain is truncated at $z=-10$ since we do not expect RTIs to form this deep as heating is negligible there. The far-field boundary conditions \eqref{eq:InfBC} are replaced with an insulating free-slip wall, 
\begin{equation}\label{eq:DNSBC}
    \frac{\partial u}{\partial z} = w = \frac{\partial T}{\partial z} = 0.
\end{equation}
This removes the need to resolve any boundary layers at the bottom boundary and maintains  the far field temperature $T_H$. The dependence of results on the grid resolution are discussed in Appendix~\ref{app:grid-res}, finding that the initial development of RTIs and kinetic energy are well resolved and insensitive to the number of points ($M$). To perform the time integration, a 3rd order, 4-stage Runge-Kutta time-stepping scheme is implemented using a fixed time step $\Delta t=2\times10^{-4}$. Every 10 time steps, Dedalus automatically verifies that $\Delta t$ satisfies a Courant-Friedrichs-Lewy-type condition to adequately resolve physical timescales \citep[see][]{Burns2020}. 

As discussed in \citet{Smith2025}, for the cryostratified regime, there is a non-trivial period ($t_0\gg1$) of gravitational stability and quiescence where the dynamics are well approximated by the heat equation \eqref{eq:heatD} with $\vec{u}=0$. This quiescent period ($t_0^*$) scales with the forcing time ($\tau_f\equiv\lambda^2 T_H/HQ_0$), but upon transformation to the convective time scale ($\tau$) via the ratio $\tau_f/\tau=\gamma\sqrt{\Ra\Pr}$, it substantially increases $t_0$ in the simulations (see table~\ref{tab:t0}). As we are primarily interested in RTI formation ($t>t_0$), computational efficiency is gained by initialising the numerical simulations with the analytic solution of the quiescent heat equation at $t=t_0$, i.e. at the time when the unstable layer begins to grow. With the general solution to the one-dimensional radiatively-forced heat equation derived in Appendix~\ref{app:analytic}, the initial condition \eqref{eq:IC} is amended to 
\begin{equation}\label{eq:DNSIC}
    T(t-t_0=0) = \left(\frac{\lambda}{H}\right)^{-\sigma}T_A\left(\frac{t_0}{\gamma\sqrt{\Ra\Pr}}, z\right) - \frac{\TMD}{T_\lambda},
\end{equation}
where 
\begin{align}\label{eq:1DforcingT0sol}
    T_A(t_f,z) = -&\frac{1}{4}\int_{-\infty}^0\left[\erf{\tilde{z}-\tilde{\zeta}} + \erf{\tilde{z}+\tilde{\zeta}}\right]\tanh\left(\frac{c_1\zeta+1}{c_2}\right)d\zeta + \frac{1}{\gamma}\erfc{-\tilde{z}} \nonumber \\[5pt] 
    &- \frac{e^z}{\gamma} + \frac{e^{\gamma t_f}}{2\gamma} \left[e^z\erfc{\sqrt{\gamma t_f}+\tilde{z}} - e^{-z}\erfc{\sqrt{\gamma t_f}-\tilde{z}}\right] - \frac{z}{2},
\end{align}
with $\tilde{z} = z/(2\sqrt{\gamma t_f})$ and $\tilde{\zeta} = \zeta/(2\sqrt{\gamma t_f})$.
This describes the temperature evolution in the quiescent regime ($t\leq t_0$) with respect to the forcing time ($t_f=t^*/\tau_f$). $T_A$ is computed using the Gauss-Kronrod 15-point rule for semi-infinite domains \citep{Virtanen2020}. To verify that the analytical solution \eqref{eq:DNSIC} is valid, the heat equation (\ref{eq:heatND} with $\vec{u}\equiv0$) is numerically solved using Dedalus \citep{Burns2020}. The semi-infinite domain is again truncated to $z\in[-10,0]$ and, at the bottom boundary, a homogenous Neumann condition ($\partial T/\partial z|_{z=-10}=0$) is employed. A time step of $10^{-3}$ is used with 256 Chebyshev collocation points. The results are shown in figure~\ref{fig:analytic-sols}, illustrating that the analytical and numerical solutions to the 1D quiescent heat equation are in close agreement. Also shown in figure~\ref{fig:analytic-sols}b is the simulation of the full convective system \eqref{eq:NS} over the quiescent period for the base configuration (table~\ref{tab:cases}). The close agreement between the 2D temperature averaged over the width $x$ and the quiescent solutions verifies the assumption that the dynamics are governed by the quiescent ($\vec{u}=0$) heat equation~\eqref{eq:heatND}.    

The time when the water column becomes gravitationally unstable ($t_0$) is defined as the first instance where $\partial T/\partial z\geq0$ at any depth. This depends on the initial thermal stratification; for example, in the cryomictic profiles the temperature gradient decays rapidly with depth, leaving the water below $z=-1$ nearly isothermal (figure~\ref{fig:ICs}a). Radiative heating reverses the weak initial gradient almost immediately, and thus $t_0 \approx 0$. Conversely, for the cryostratified profiles an appreciable stabilising temperature gradient persists at $z=-1$, and a finite period of radiative heating is required to overcome it. However, in the hypolimnion we find that weak positive temperature gradients develop from trace amounts of heating due to the initially zero gradient. Since we are only concerned with the onset of an appreciable instability within the thermocline, for the cryostratified cases, we filter these arbitrarily small instabilities with the criterion of $|\rho(t,z)-\rho(t_0,z)|/(\rhoMD-\rho_0)\geq10^{-3}$ when computing $t_0$ from the simulations. These calculated values of $t_0$ are listed in table~\ref{tab:t0} expressed in both the forcing ($t^*/\tau_f$) and convective ($t^*/\tau$) timescales.  

\begin{table}
\centering
\setlength{\tabcolsep}{8pt}
\begin{tabular}{c c c c}
 & \multicolumn{2}{c}{Cryostratified} & \citet{Smith2025} \\
\cmidrule(lr){2-3} \cmidrule(lr){4-4}
Configuration & $t_0^*/\tau$ & $t_0^*/\tau_f$ & $t_0^*/\tau_f \approx e^{5.3\sqrt{\gamma}}$ \\
\midrule
Baseline & 53.253 & 1.684 & 1.699 \\
High $\gamma$ & 3231.85 & 10.22 & 5.344 \\
Low $\gamma$ & 3.763 & 1.19 & 1.182 \\
Low $\Ra$ & 16.84 & 1.684 & 1.699 \\
High $\Ra$ & 168.4 & 1.684 & 1.699 \\
Low $\delta/H$ & 53.253 & 1.684 & 1.699 \\
High $\delta/H$ & 52.525 & 1.661 & 1.699 \\ 
\end{tabular}
\caption{\label{tab:t0} Onset of gravitational instability, given in terms of convective ($t_0^*/\tau$) and forcing ($t_0^*/\tau_f$) time scales across all the cryostratified configurations defined in Table~\ref{tab:cases} and compared against the small-$\gamma$ approximation by \citet{Smith2025}.}
\end{table}

Dedalus implements high order spectral schemes which feature very low levels of numerical noise and thus instability development is suppressed compared to experimental or field conditions. To allow for RTI growth, artificial noise is manually embedded into the initial temperature distribution \eqref{eq:DNSIC}. Similar approaches have been adopted in other studies and the added fluctuations can be interpreted physically as representing a range of mechanisms, such as wind-driven oscillations under the ice cover \citep[e.g.][]{cortes2020, Castendyk2021} or small swimming organisms \citep[e.g.][]{noss2012}. Following \citet{Smith2025}, small initial perturbations of the form $\tilde{T}_0/T_\lambda$ are added to the initial state \eqref{eq:DNSIC}, where
\begin{equation}\label{eq:initial_perts}
    \tilde{T}_0 = \frac{\tilde{S}\theta}{\max|\tilde{T}_r|}\tilde{T}_r(x,z)\sin\left(\frac{\pi|z|}{10}\right).
\end{equation}
Here, random fluctuations ($\tilde{T}_r$) are obtained using a sine expansion in both $x$ and $z$ of the form
\begin{equation}\label{eq:noise-structure}
    \tilde{T}_r(x,z) = \sum_{m=0}^{200}\sum_{n=0}^{200} c_{m,n}\sin\left(\frac{m\pi x}{20}\right)\sin\left(\frac{n\pi z}{10}\right),
\end{equation}
where the $201\times201$ random coefficients ($c_{m,n}$) are drawn once from a uniform distribution between $\pm0.01$ and are utilised for all simulations to ensure a reproducible structure. Noise is damped at both walls by multiplying by $\sin(\pi|z|/10)$ before normalising by $\max{|\tilde{T}_r|}$ and rescaling by $\tilde{S}\theta$ where $\theta$ is the perturbation temperature scale (Appendix~\ref{app:LA}) and $\tilde{S}=10^{-2}$ is a noise amplitude. The ratio of the background to perturbation temperature scales can be written as $T_\lambda/\theta=\gamma\sqrt{\Ra\Pr}$. This formulation is advantageous as the initial perturbations remain consistent throughout. Results were found to be generally insensitive to variations to $\tilde{T}_r$ but somewhat dependent on the magnitude $\tilde{S}$ (Appendix~\ref{app:noise}). 

%%%%%%%%%%%%%%%%%%%%%%%%%%%%%%%%%%%%%%%%%%%%%%%%%%%%%%%%%%%%%%%%
%%%%%%%%%%%%%%%%%%%%%%%%%%%%%%%%%%%%%%%%%%%%%%%%%%%%%%%%%%%%%%%%
\section{Simulation Results}
\label{sec:dns-results}
%%%%%%%%%%%%%%%%%%%%%%%%%%%%%%%%%%%%%%%%%%%%%%%%%%%%%%%%%%%%%%%%

\subsection{Unstable Layer Formation ($t\leq t_0$)}

\begin{figure}
    \centering
    \includegraphics[width=\linewidth]{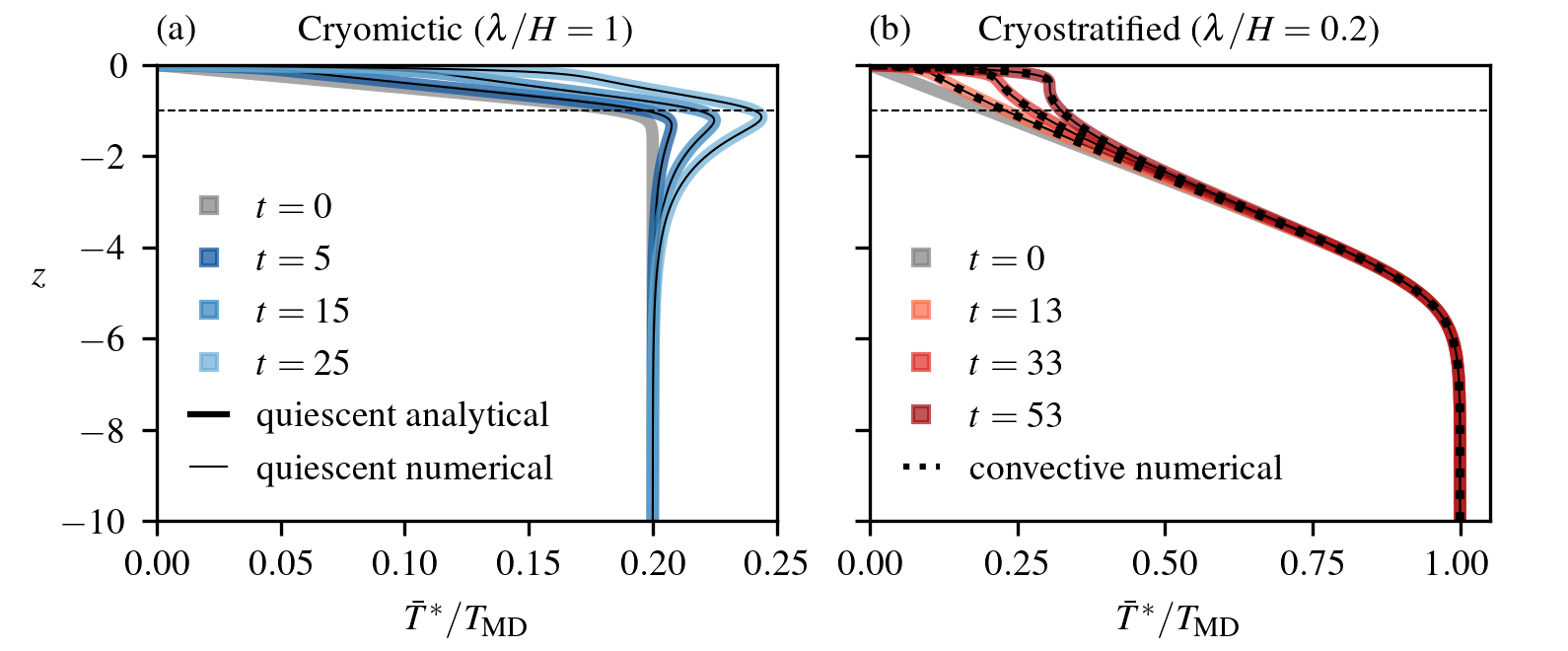}
    \caption{\label{fig:analytic-sols} Temperature profiles for $\gamma=0.01$ and $\Ra=10^6$ at different times for the cryomictic (a), and cryostratified (b) regimes. Thin dashed lines indicate $z^*=-\lambda$ where there is substantial light absorption and heating. Analytical (equation~\ref{eq:1DforcingT0sol}, solid coloured lines) and numerical solutions (thin solid black lines) to the quiescent heat equation (\ref{eq:heatND} with $\vec{u}\equiv0$) show excellent agreement. Full numerical solutions to \eqref{eq:NS} averaged over $x$ (equation~\ref{eq:average}, dotted lines) for the cryostratified regime also agree well until $t=t_0\approx53$.}
\end{figure}

The formation of the gravitationally unstable layer for the cryomictic and cryostratified profiles is illustrated in figure~\ref{fig:analytic-sols}. We show the time evolution ($t\leq t_0$) of the temperature profiles for $\gamma=0.01$, $\Ra=10^6$ and $\delta/H=0.2$. For both regimes, heat is deposited in the upper water column which destroys the monotonicity of the temperature, and therefore density, and leads to gravitational instability. 

For the cryomictic case (figure~\ref{fig:analytic-sols}a), the water column becomes unstable almost immediately as substantial heat is added to the hypolimnion ($z<-1$). For the cryostratified case (figure~\ref{fig:analytic-sols}b), the unstable layer forms in the thermocline ($z>-1$) and requires heating over an extended period to develop instability by $t_0\approx53$ (table~\ref{tab:t0}). 

Values of $t_0$ for the cryostratified configurations are compared in table~\ref{tab:t0} to the small $\gamma$ approximation derived by \citet{Smith2025} for a linear thermocline. There is overall close agreement except for the high $\gamma$ simulation where $\gamma$ is increased to 0.1 (violating the small-$\gamma$ approximation). Agreement slightly decreases for the high $\delta/H$ configuration where the larger $\delta/H=0.4$ reduces the linearity of the thermocline. 

\subsection{RTI Formation ($t\geq t_0$)}

\begin{figure}
    \centering
    \includegraphics[width=\linewidth]{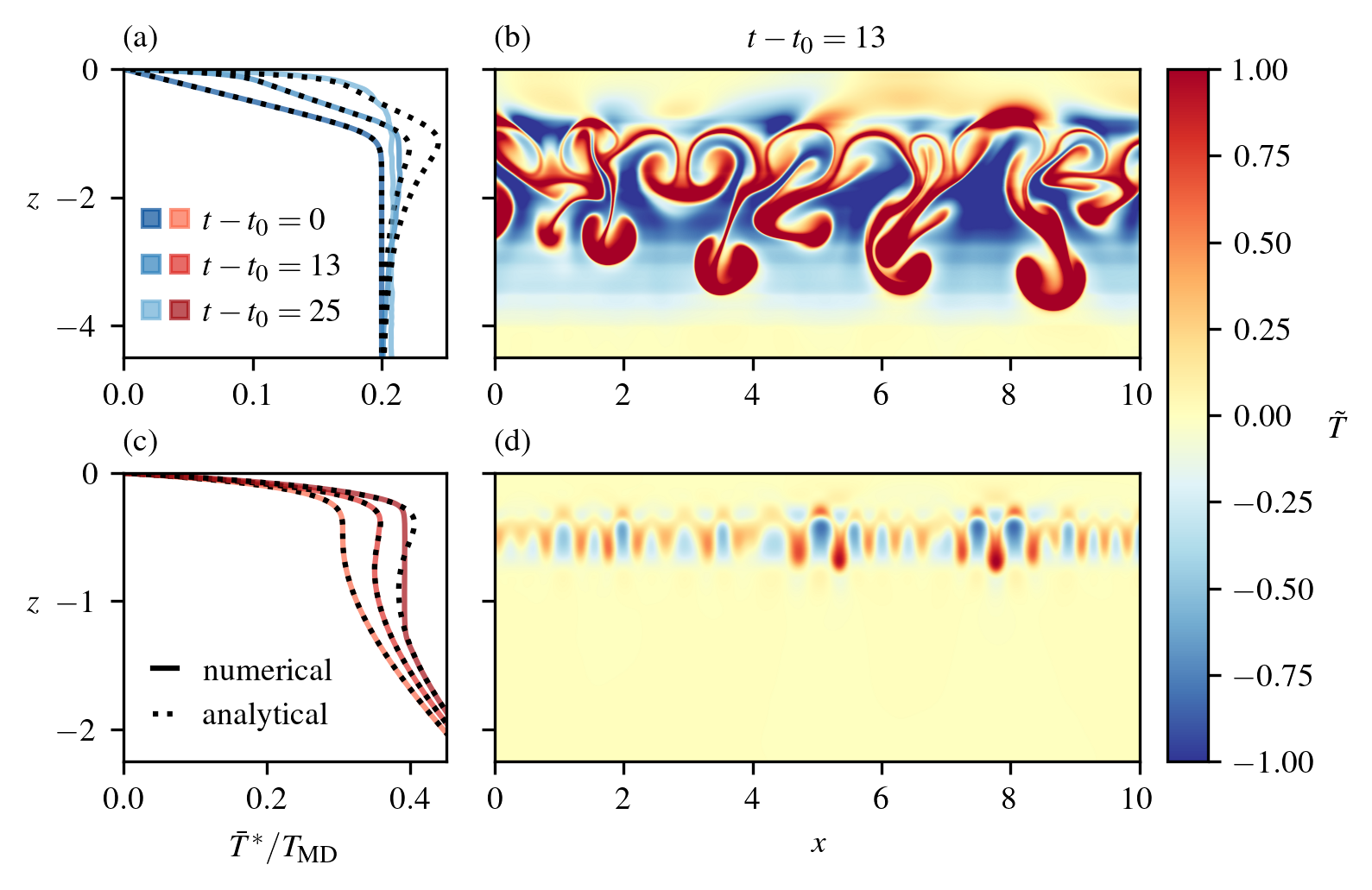}
    \caption{\label{fig:DNS} Two-dimensional DNS results for $\Ra=10^6$ and $\gamma=0.01$. The top row corresponds to the cryomictic regime ($\lambda/H=1$, $T_H/\TMD=0.2$) and the bottom row for the cryostratified regime ($\lambda/H=0.2$, $T_H/\TMD=1$). The left column (a, c) displays the evolution of the span averaged temperature profile scaled by the temperature of maximum density $\TMD$ for the simulations (solid lines) against the analytical solution (equation~\ref{eq:1DforcingT0sol}, dotted lines). The right column (b, d) displays a snapshot of the temperature perturbations ($\tilde{T}$) at time $t-t_0=13$.}
\end{figure}

Once the temperature profile becomes gravitationally unstable ($t=t_0$), small perturbations grow and lead to RTI formation. The convective mixing subsequently drives deviation from the analytic quiescent solution (figure~\ref{fig:DNS}a). The evolution of the horizontally averaged temperature profile is plotted for $\gamma=0.01$ and $\Ra=10^6$ (figure~\ref{fig:DNS}a, c), defined as 
\begin{equation}\label{eq:average}
    \bar{T}(z,t) = \frac{1}{X}\int_0^{X} T(x,z,t)\; dx,
\end{equation}
where $X = 20$ is the horizontal width of the domain. To aid comparison, the temperature is rescaled by $\TMD$, specifically, $\bar{T}^*/\TMD = T_\lambda\bar{T}/\TMD + 1$ (see equation \ref{eq:scalings}). In the right column of figure~\ref{fig:DNS}, snapshots of the temperature perturbations are shown which are obtained via $\tilde{T}=T_\lambda(T-\bar{T})/\theta=\gamma\sqrt{\Ra\Pr}(T-\bar{T})$. 

For both cases, RTIs have formed by $t-t_0=13$ but the cryomictic simulation (figure~\ref{fig:DNS}b) is at a more advanced stage compared with the cryostratified simulation (figure~\ref{fig:DNS}d). This suggests that RTIs form quicker for cryomictic lakes than for cryostratified lakes even when the development of the unstable layer ($t_0$) is removed. This may be due to RTIs fighting against the stable stratification of the thermocline whereas the isothermal hypolimnion allows for quicker growth. Additionally, the depth of RTI formation is co-located with the depth of the unstable layer: in figures~\ref{fig:DNS}b (cryomictic) and \ref{fig:DNS}d (cryostratified) the location of where RTIs originate approximately matches the depth where the water column first becomes unstable, at $z\approx-1$ and $z\approx-0.3$ respectively. This behaviour is robust for all simulations that develop RTIs. Additionally, RTI growth in the hypolimnion is not observed for the cryostratified simulations since the available residual forcing in the hypolimnion is too weak for RTIs to develop. Once RTIs have formed ($t-t_0>13$), differences between the analytical and numerical solutions continue to increase. By $t-t_0=25$, a mixing layer has clearly developed characterised by the near-zero gradient of the temperature profiles before it transitions back into the thermocline. 

\begin{figure}
    \centering
    \includegraphics[width=\linewidth]{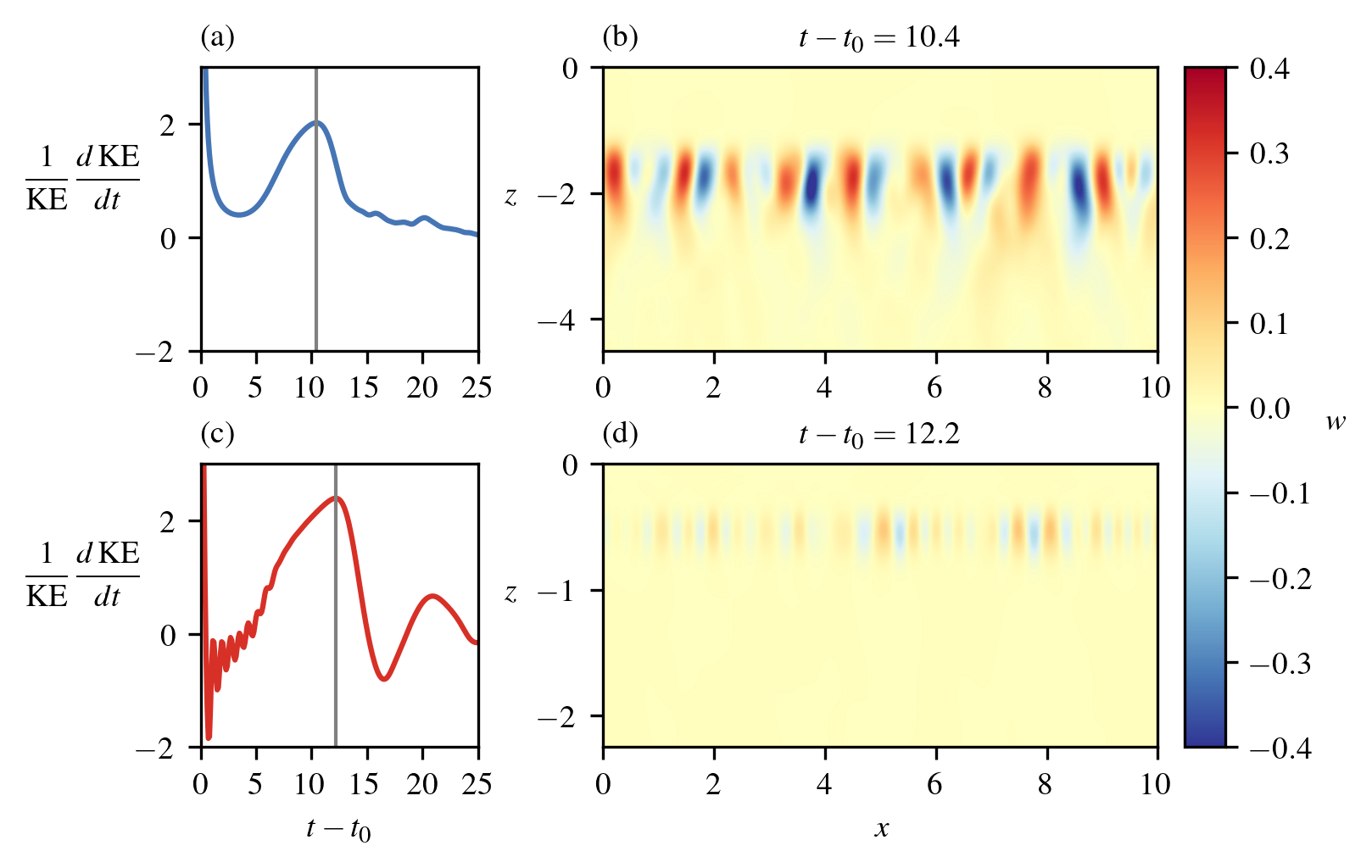}
    \caption{\label{fig:KE-t1} Two-dimensional DNS results for $\Ra=10^6$ and $\gamma=0.01$. The top row corresponds to the cryomictic regime ($\lambda/H=1$, $T_H/\TMD=0.2$) and the bottom row to the cryostratified regime ($\lambda/H=0.2$, $T_H/\TMD=1$) initiated at $t_0$. The left column (a, c) displays the evolution of the KE growth rate. The right column (b, d) displays snapshots of the velocity perturbations ($w$) at the time of peak KE growth (vertical grey lines in b, c).}
\end{figure}

We refer to $t = t_1$ as the time of convection onset, or the time in which nonlinear growth of RTIs initiate large-scale convective plumes in the flow-field. Thus, we define $t_1$ as the time of peak growth in the volume-averaged kinetic energy (KE),
\begin{equation}\label{eq:KEgrowth}
    \text{KE} = \frac{1}{2V}\int_V u^2+w^2\; dV, 
\end{equation}
over the domain $V$. In figure~\ref{fig:KE-t1}, the relative growth rate of kinetic energy, which we define as $d \ln(\text{KE}) /dt  = (d\text{KE}/dt)/\text{KE}$, is plotted. This measure is chosen by assuming that the kinetic energy growth is exponential (i.e. $\text{KE}\sim e^{g(t)}$), then the natural logarithm yields the exponent (i.e. $\ln(\text{KE})\sim g(t)$) and $d \ln(\text{KE})/dt \sim dg/dt$ gives the growth rate. 

\begin{figure}
    \centering
    \includegraphics[width=\linewidth]{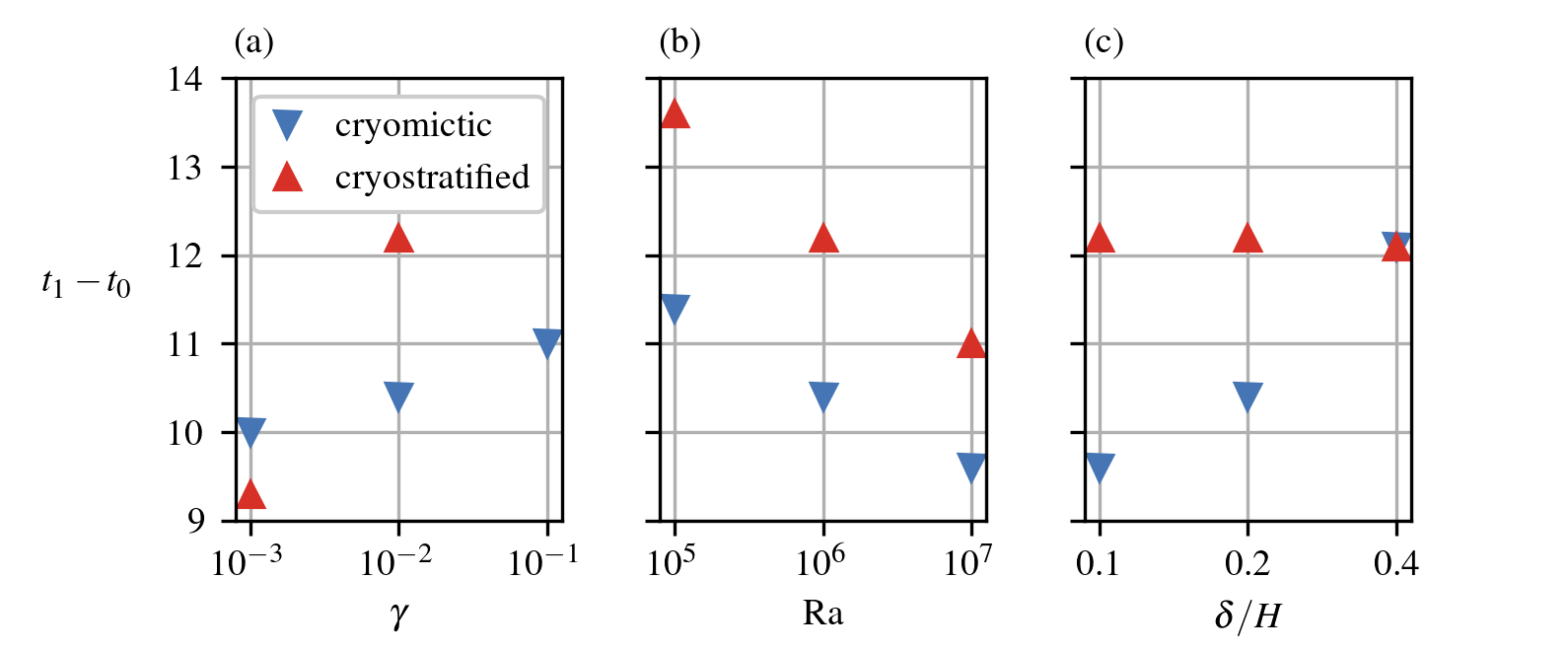}
    \caption{\label{fig:t1} Timing of convection onset, $t_1-t_0$, for (a) varying $\gamma$ with fixed $\Ra=10^6$ and $\delta/H=0.2$, (b) varying $\Ra$ with fixed $\gamma=10^{-2}$ and $\delta/H=0.2$, and (c) varying $\delta/H$ with fixed $\gamma=10^6$ and $\Ra=10^6$. Note, there is no data point for the cryostratified $\gamma=0.1$ configuration since convection was not observed over the time span of the simulations ($t-t_0\leq25$).}
\end{figure}

Clearly defined peaks are observed in figures~\ref{fig:KE-t1}a and c. These are interpreted as marking a physical change in behaviour, where the kinetic energy transitions from a period of growth to a period of decay. We argue that this represents a transition from the linear growth stage into the nonlinear regime, where plumes begin to form inducing shear and rotation allowing the characteristic RTI shape to emerge. These signatures are cross-referenced against snapshots of the vertical velocity ($w$) in figures~\ref{fig:KE-t1}b and d where clear signatures of RTI formation are seen. This is robust for all cases, except for the $\gamma=0.1$ cryostratified simulation where convection is not observed over the time period of the simulation. As a result, we interpret the timing of the KE growth peaks as the onset of convection ($t_1$). These are plotted, relative to $t_0$, in figure~\ref{fig:t1} and it is generally seen that convection begins earlier for the cryomictic regime compared to the cryostratified regime.  

As $\gamma$ is decreased (figure~\ref{fig:t1}a) for fixed $\Ra=10^6$ and $\delta/H=0.2$, the time to convection decreases due to the increase in forcing. For $\gamma=10^{-3}$, there is quicker onset for the cryostratified configuration. This suggests a transition point where there is enough forcing to overcome the stabilising properties of the thermocline and to even initiate convection before the cryomictic configuration (albeit relative to $t_0$ which we expect would be much larger).  

As $\Ra$ is increased (figure~\ref{fig:t1}b) with $\gamma=10^{-2}$ and $\delta/H=0.2$ fixed, the time to RTI formation decreases. Physically, we interpret that an increase in $\Ra$ acts to speed up RTI development due to the suppression of diffusive, relative to buoyant, processes.

Lastly, $\delta/H$ is varied with fixed $\gamma=0.01$ and $\Ra=10^6$ (figure~\ref{fig:t1}c), to isolate the role of the abruptness of the thermocline--hypolimnion transition. It is observed that changing $\delta/H$ has little-to-no impact for cryostratified configurations because RTI development is confined to the linear thermocline where curvature of the temperature profile is minimal. For the cryomictic cases, $\delta/H$ is more influential, delaying RTI formation with increasing $\delta/H$. This is because RTIs form exactly where the thermocline transitions into the hypolimnion meaning that curvature effects in the water temperature will amplify and influence the dynamics.  

It is interesting to note that for all configurations (apart from the $\gamma=0.1$ simulation) $t^*_1-t^*_0\sim O(10\tau)$ despite the large parameter range considered. This suggests that the more variable period of unstable layer formation, ranging from $t_0 = 4$ to $3200$ in our simulations, is the dominant factor for the timing of RTIs, and the delay of convection beyond $t_0$ is relatively insensitive.  

%%%%%%%%%%%%%%%%%%%%%%%%%%%%%%%%%%%%%%%%%%%%%%%%%%%%%%%%%%%%%%%%
%%%%%%%%%%%%%%%%%%%%%%%%%%%%%%%%%%%%%%%%%%%%%%%%%%%%%%%%%%%%%%%%
\section{Linear Growth of RTIs ($t_0< t< t_1$)}
\label{sec:linear}
%%%%%%%%%%%%%%%%%%%%%%%%%%%%%%%%%%%%%%%%%%%%%%%%%%%%%%%%%%%%%%%%

As seen in the previous section, once the water column becomes gravitationally unstable ($t> t_0$) small perturbations can grow to form RTIs and initiate large scale convective plumes at $t_1$. To build understanding, we assume that during the early growth stage before RTI formation ($t_0<t<t_1$), the amplitudes of the perturbations remain sufficiently small so that nonlinear effects are negligible \citep{chang2023,yang2026}, and the dynamics can be well predicted by linear analysis. 

\subsection{Linear Analysis}

To linearise the system \eqref{eq:NSD}, the flow variables are decomposed into background quantities (bars) and small perturbations (tildes) in Appendix~\ref{app:LA-pert-eqns}. The linearised system is scaled
%\eqref{eq:lin_NSD} 
similarly as in section~\ref{sec:scaling}, but the temperature is scaled by $T^*=T_\lambda\bar{T}+\TMD+\theta\tilde{{T}}$ to give
\begin{subequations}\label{eq:lin_NS}
    \begin{gather}
        \bnabla\bcdot\tilde{\vec{u}} = 0, \\[5pt]
        \frac{\partial\tilde{\vec{u}}}{\partial t}
        = -\bnabla\tilde{P}
        +\sqrt{\frac{\Pr}{\Ra}}\nabla^2\tilde{\vec{u}}
        + \beta\bar{T}\tilde{T}\vec{\hat{e}}_z, \\[5pt]
        \frac{\partial\tilde{T}}{\partial t}
        + \left(\frac{\lambda}{H}\right)^{\sigma-1}\gamma\sqrt{\Ra\Pr}\,\tilde{w}\frac{\partial\bar{T}}{\partial z}
        = \frac{1}{\sqrt{\Ra\Pr}}\nabla^2\tilde{T}, \label{eq:lin_NS_heat}
    \end{gather}
\end{subequations}
subject to the homogeneous boundary conditions,
\begin{equation}\label{eq:NBCs}
\tilde{u} = \tilde{w} = \tilde{T} = 0 \;\text{at } z=0 \text{ and as } z \to -\infty.
\end{equation}

Since the background evolves with time, solutions are assumed in the form of horizontal Fourier modes, 
\begin{equation}\label{eq:fourier}
[\tilde{u}, \tilde{w}, \tilde{P}, \tilde{T}]^\text{T} = \left[\hat{u}(z,t),\hat{w}(z,t),\hat{P}(z,t),\hat{T}(z,t)\right]^\text{T}e^{ikx},
\end{equation}
where $k$ is the wavenumber. After substitution into \eqref{eq:lin_NS},
\begin{subequations}\label{eq:2D_pert_eqns}
    \begin{gather}\label{eq:2D_pert_eqns_cont}
        ik\hat{u} + \frac{\partial \hat{w}}{\partial z} = 0, \\[4pt]
        \frac{\partial\hat{u}}{\partial t} = -ik\hat{P} + \sqrt{\frac{\Pr}{\Ra}}\left(\frac{\partial^2 \hat{u}}{\partial z^2} - k^2 \hat{u}\right), \\[4pt]
        \frac{\partial \hat{w}}{\partial t} = -\frac{\partial \hat{P}}{\partial z} + \sqrt{\frac{\Pr}{\Ra}}\left(\frac{\partial^2 \hat{w}}{\partial z^2} - k^2 \hat{w}\right) +\beta\bar{T}\hat{T}, \\[4pt]
        \frac{\partial \hat{T}}{\partial t} + \left(\frac{\lambda}{H}\right)^{\sigma-1}\gamma\sqrt{\Ra\Pr}\,\hat{w}\frac{\partial \bar{T}}{\partial z} = \frac{1}{\sqrt{\Ra\Pr}}\left(\frac{\partial^2 \hat{T}}{\partial z^2} - k^2 \hat{T}\right),
    \end{gather}
\end{subequations}
with the boundary conditions 
\begin{equation}\label{eq:2D_perts_BCs}
    \hat{u} = \hat{w} = \hat{T} = \frac{\partial\hat{w}}{\partial z} = 0 \; \text{at } z=0 \text{ and as } z \to -\infty,
\end{equation}
where the latter condition is a result of the incompressibility condition \eqref{eq:2D_pert_eqns_cont}. 

To solve the peturbation equations \eqref{eq:2D_pert_eqns}, we apply a Chebyshev collocation scheme. In summary, it expresses the solution as a truncated series of Chebyshev polynomials and transforms \eqref{eq:2D_pert_eqns} into a matrix system for the coefficients ($a_s$) of the Chebyshev polynomials. This approach has been widely adopted in flow stability \citep[e.g.][]{schmid2001,christopher2023} due to its quick convergence and natural clustering of points near boundaries that aid in resolving boundary layers. A null space method, similarly to \citet{christopher2023}, is applied that expresses the coefficients ($a_s$) in terms of the null space formed by the homogenous boundary conditions \eqref{eq:2D_perts_BCs} and continuity equation \eqref{eq:2D_pert_eqns_cont}. The perturbation equations \eqref{eq:2D_pert_eqns} can then be concisely written as 
\begin{equation}\label{eq:bstprime}
    \frac{d\vec{b}_s}{dt} = \mat{Q}(t)\,\vec{b}_{s},
\end{equation} 
where $\vec{b}_s$ are the coordinates of the null space, or equivalently, through Chebyshev expansions, coordinates of each mode (i.e. $\hat{u}$, $\hat{w}$, $\hat{T}$, $\hat{P}$). Full details, including the construction of the matrix $\mat{Q}(t)$, are provided in Appendix~\ref{app:LA-methods} and supplementary material S2. 

For each wavenumber $k$, the linearity of \eqref{eq:2D_pert_eqns} allows the state at time $t$ to be related to its initial state at $t_0$ through a finite-time propagator $\mat{M}$ \citep{linot2024},
\begin{equation}\label{eq:finite_time_propagator}
    \vec{b}_{s}(t;k) = \mat{M}(t;k)\vec{b}_{s}(t_0).
\end{equation}
$\mat{M}$ is approximated numerically by partitioning the interval $[t_0,t]$ into $\tilde{M}$ subintervals with endpoints $t_0=\tau_0<\tau_1<\cdots<\tau_{\tilde{{M}}}=t$. On each subinterval, $\mat{Q}(t;k)$ is approximated by its value at the midpoint and upon integrating
\begin{equation}\label{eq:matrix_integration}
    \vec{b}_s(\tau_m;k)\approx\mat{E}_m(k)\vec{b}_s(\tau_{m-1};k),
    \qquad
    \mat{E}_m(k)=\exp\left[
    \Delta\tau_m,\mat{Q}\left(\frac{\tau_m+\tau_{m-1}}{2};k\right)
    \right],
\end{equation}
where we set $\Delta\tau_m=\tau_m-\tau_{m-1}=0.1$. Applying this integration successively yields
\begin{equation}\label{eq:matrix_integration_multiplication}
    \vec{b}_s(t;k)\approx
    \mat{E}_{\tilde{M}}(k)\mat{E}_{\tilde{M}-1}(k)\cdots\mat{E}_1(k)\vec{b}_s(t_0;k),
\end{equation}
so that the propagator satisfies
\begin{equation}\label{eq:propagator_product}
    \mat{M}(t;k) \approx \mat{E}_{\tilde{M}}\mat{E}_{\tilde{M}-1} \cdots \mat{E}_{1}.
\end{equation}

\subsection{Amplification Theory}

Here, the metrics used to determine the amplification, or gain/growth, of perturbations are outlined. For a given wavenumber $k$, we define the modal amplification factor as the spectral radius of the finite-time propagator $\mat{M}$, or equivalently the maximum absolute valued eigenvalue $\mu_j$ of $\mat{M}(t;k)$, i.e.
\begin{equation}\label{eq:Gtk}
    G_L(k,t) = \max_j |\mu_j|.
\end{equation}
This measures the largest amplification attainable by an individual eigenmode. It does not, in general, bound the amplification of an arbitrary initial perturbation which depends on the measure used, such as the $L_2$ norm or equation~\eqref{eq:mechE} below, to measure the amplitude and can exceed $G_L$. Instead, focus is on the modal amplification because of the observation that perturbations in the simulations are predominantly governed by a single mode (figures~\ref{fig:KE-t1}b, d), and our objective is to identify the dominant finite-time instability branch and quantify its growth.  

To compare to the numerical simulations (section~\ref{sec:dns-results}), we define and compute the modal mechanical energy of the numerical output as
\begin{equation}\label{eq:mechE}
    \hat{E}(k,t) = \frac{1}{2Z}\int_{-Z}^0 |\hat{u}|^2 + |\hat{w}|^2 \; dz - \frac{\beta}{2Z}\frac{\theta}{T_\lambda}\int_{-Z}^0 z|\hat{T}|^2 \; dz,
\end{equation}
over the depth $Z=10$ and where the modes ($\hat{u}$, $\hat{w}$, $\hat{T}$) are computed using Fast Fourier Transforms \citep{Virtanen2020}. Adapted from \citet{winters2019}, the first and second integrals denote the depth averaged kinetic and potential energies of each mode, respectively. Here, the potential energy is included as to remove the singularity ($\text{KE}=0$) present at $t-t_0=0$ due to the quiescent initial condition. Consequently, the amplification or gain of energy for each mode can be defined as
\begin{equation}\label{eq:Egrowth}
    G_E(k,t) = \sqrt{\frac{\hat{E}(k,t)}{\hat{E}(k,t_0)}}.
\end{equation}

Finally, the most amplified wavenumber is defined as the wavenumber that attains the maximum amplification over all wavenumbers, such that
\begin{equation}
    G_E(k_{\max},t) = \max\limits_{k} G_E(k,t) \quad\text{and}\quad G_L(\tilde{k}_{\max},t) = \max\limits_{k} G_L(k,t).
\end{equation}
Here, $k_{\max}$ refers to the most amplified, or unstable, wavenumber from the simulations using equation~\eqref{eq:Egrowth}, and $\tilde{k}_{\max}$ refers to the most amplified wavenumber predicted by the linear analysis using equation~\eqref{eq:Gtk}. 

\subsection{Results}

\begin{figure}
    \centering
    \includegraphics[width=\linewidth]{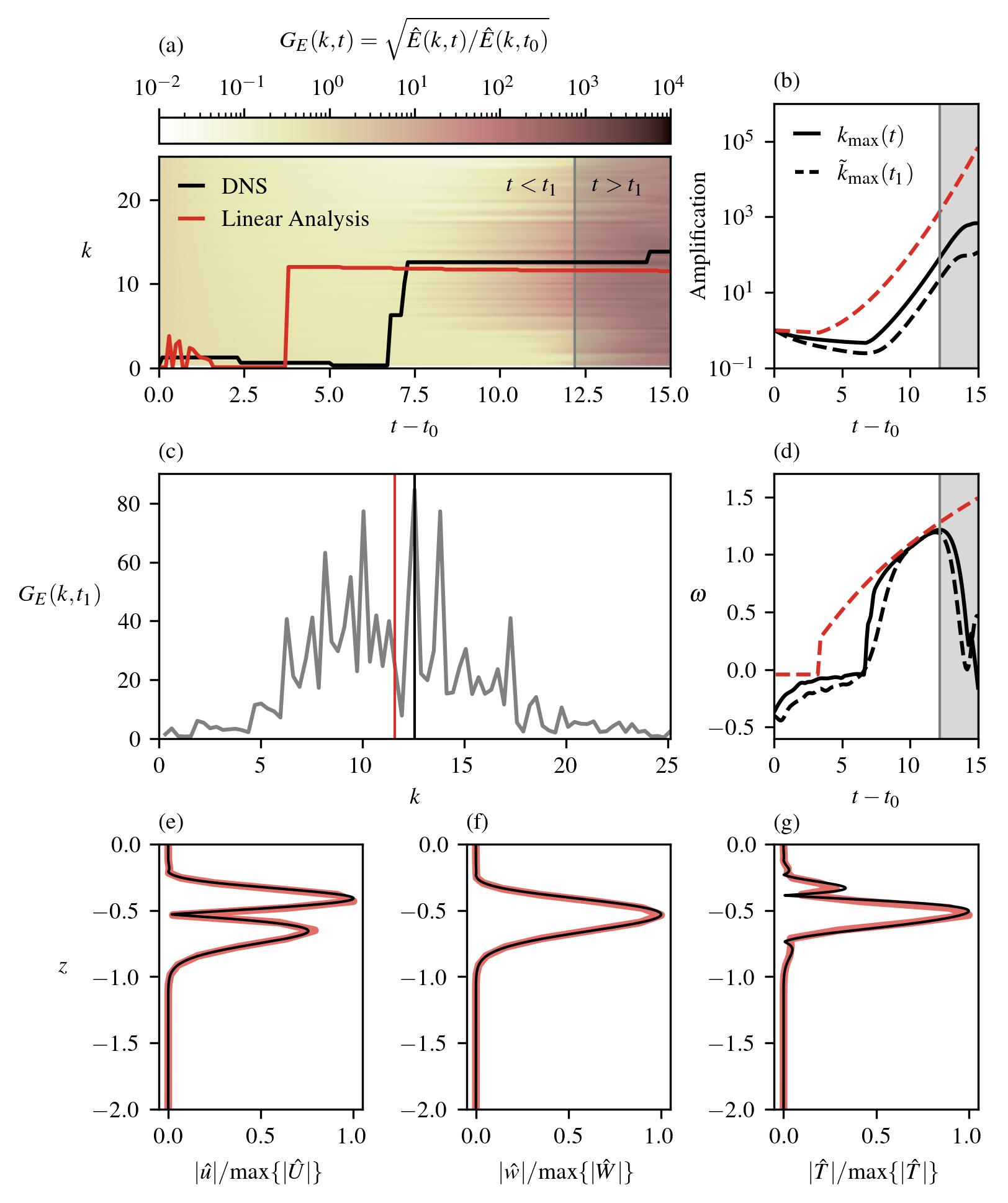}
    \caption{\label{fig:csrun1-compare} Comparison of the linear analysis to the numerical simulations for the cryostratified regime ($\lambda/H=0.2$, $T_H/\TMD=1$) with $\Ra=10^6$ and $\gamma=0.01$. (a) Spectrum of modal energy gain ($G_E$) and the time evolution of the most amplified/unstable wavenumber as predicted by the linear analysis ($\tilde{k}_{\max}$, red) compared to the simulations ($k_{\max}$, black). Grey shaded boxes denote times after $t_1$. (b) Evolution of the amplification/gain predicted by linear analysis ($G_L$, red) compared to the simulations ($G_E$, black) for the time-varying observed most unstable wavenumber ($k_{\max}$, solid line) and predicted most unstable wavenumber at $t_1$ ($\tilde{k}_{\max}$, dashed lines). (c) Spectrum of gain at time $t_1-t_0=12.2$ (grey lines) with the observed most unstable wavenumber ($k_{\max}$, black line) and predicted most unstable wavenumber ($\tilde{k}_{\max}$, red line). (d) Time evolution of the relative growth rate ($\omega$) comparing the predicted (linear analysis, red) to the observed (DNS, black) for $k_{\max}$ (solid) and $\tilde{k}_{\max}(t_1)$ (dashed). (e), (f), (g) Comparison of the modal structures predicted by linear analysis (red) and observed by DNS (black) for $\tilde{k}_{\max}$ at time $t_1-t_0=12.2$.}
\end{figure}

\begin{figure}
    \centering
    \includegraphics[width=\linewidth]{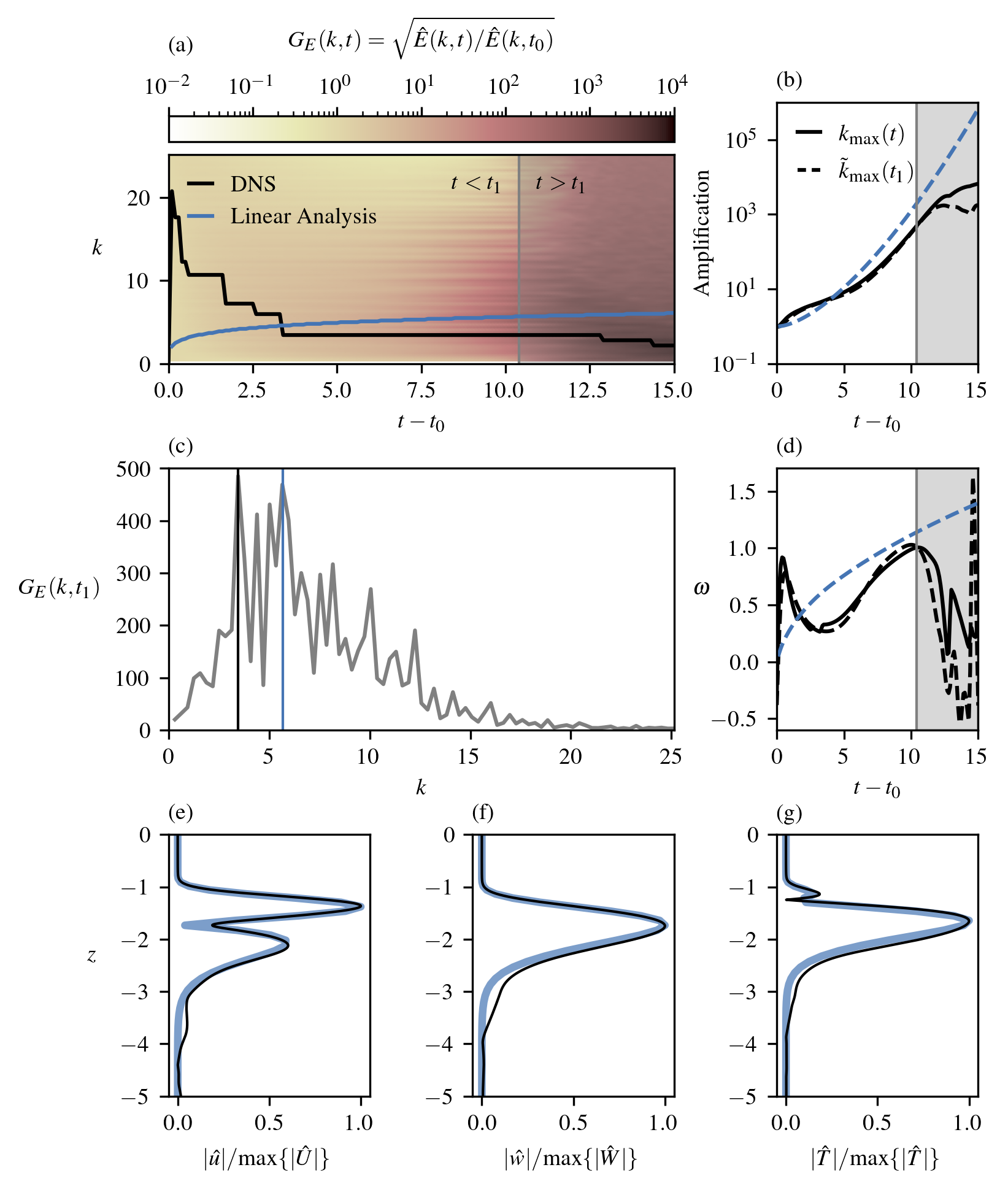}
    \caption{\label{fig:cmrun1-compare} Same as figure~\ref{fig:csrun1-compare} but for the cryomictic regime ($\lambda/H=1$, $T_H/\TMD=0.2$) with $\Ra=10^6$ and $\gamma=0.01$. Note snapshots taken from $t_1-t_0=10.4$ for (c), (e), (f) and (g).}
\end{figure}

The early RTI growth dynamics are estimated using equation~\eqref{eq:propagator_product} with 128 Chebyshev polynomials until a final time $\tau_{\tilde{M}}=15$. Similarly to the simulations, the domain is truncated at a depth $Z=10$, and we find that results are not sensitive to variations in $\tau_{\tilde{M}}$ and $Z$ or to the different boundary conditions implemented for the simulations \eqref{eq:DNSBC} and linear analysis \eqref{eq:2D_perts_BCs} (not shown). In figures~\ref{fig:csrun1-compare} and \ref{fig:cmrun1-compare}, results are plotted for the cryostratified ($\lambda/H=0.2$, $T_H/\TMD=1$) and cryomictic conditions ($\lambda/H=1$, $T_H/\TMD=0.2$), respectively, at fixed $\Ra=10^6$ and $\gamma=0.01$. 

The amplification, or gain, of modal energy ($G_E$) is plotted in figures~\ref{fig:csrun1-compare}a and \ref{fig:cmrun1-compare}a with the time evolution of the most amplified wavenumbers overlaid for the simulations ($k_{\max}$) and linear analysis ($\tilde{k}_{\max}$). The gain is also plotted for the most amplified modes as predicted by the linear analysis ($G_L$) and output by the DNS ($G_E$) in figures~\ref{fig:csrun1-compare}b and \ref{fig:cmrun1-compare}b. For the cryostratified regime, there is a period of decay (figure~\ref{fig:csrun1-compare}b) where the amplification remains small until modes begin to grow at $t-t_0\sim7.5$ signalling the onset of instability growth and highlighted by the jump in the most unstable, or most amplified, wavenumber to $k_{\max}\sim12$ (figure~\ref{fig:csrun1-compare}a). These unstable modes continue to amplify until convection is observed at $t_1-t_0=12.2$. Conversely, for the cryomictic regime, modes amplify immediately (figure~\ref{fig:cmrun1-compare}b) as the most unstable wavenumber spikes at $k_{\max}\sim 20$ before reducing to $k_{\max}\approx4$ (figure~\ref{fig:cmrun1-compare}a). These more amplified modes continue to grow until RTIs form at $t_1-t_0=10.4$. In both cases, the predictions of the most amplified wavenumber ($\tilde{k}_{\max}$) generally behave similar to the DNS, capturing both the lack of amplification in the cryostratified regime (figures~\ref{fig:csrun1-compare}a, b) and instant gain in the cryomictic regime (figures~\ref{fig:cmrun1-compare}a, b). However, the linear analysis predicts modal amplification earlier than what was observed for the cryostratified regime and, for both cases, there are slight differences between the predicted and observed most unstable wavenumbers. This latter point is most easily seen in figures~\ref{fig:csrun1-compare}c and \ref{fig:cmrun1-compare}c where the spectrum of modal energy amplification ($G_E$) at time $t_1$ is plotted. Comparing the predictions ($\tilde{k}_{\max}$) to DNS results ($k_{\max}$), there is fairly close agreement in the amplified spectrum. As discussed in Appendix~\ref{app:noise}, the most unstable wavenumber from the simulations is sensitive to the initial noise profile (figure~\ref{fig:noise-coeffs}) since it depends on how amplified each unstable mode initially is. 

The growth dynamics can be seen in figures~\ref{fig:csrun1-compare}b and \ref{fig:cmrun1-compare}b, where the amplification of the most unstable mode at each sample time in the simulations, $G_E(k_{\max},t)$, is compared to that of the predicted most unstable mode at $t=t_1$, $G_L(\tilde{k}_{\max}(t_1),t)$. Despite the earlier gain predicted by the linear analysis for the cryostratified regime (figure~\ref{fig:csrun1-compare}b), the amplification curves after the onset of instability growth (at $t-t_0\approx7.5$ where perturbations begin to amplify) are close to parallel. This suggests that the rate of amplification is consistent across both the predictions and simulations. The modal energy gain of $\tilde{k}_{\max}$, $G_E(\tilde{k}_{\max}(t_1), t)$, is plotted to directly compare to the linear analysis where there is similar behaviour as above. We choose $\tilde{k}_{\max}(t_1)$ as the representative most unstable wavenumber to reduce computational cost and since $\tilde{k}_{\max}(t)$ varies only slightly after the onset of instability growth at $t-t_0\approx4$ (figures~\ref{fig:csrun1-compare}a and \ref{fig:cmrun1-compare}a). Further, there is no noticeable difference with the amplification history ($G_L$) between $\tilde{k}_{\max}(t)$ or $\tilde{k}_{\max}(t_1)$ (not shown).   

To further investigate the agreement between the linear analysis and DNS in figures~\ref{fig:csrun1-compare}b and \ref{fig:cmrun1-compare}b, the relative growth rate is computed by
\begin{equation}\label{eq:omega}
    \omega \equiv \frac{d}{dt}\ln(G_i) = \frac{1}{G_i}\frac{dG_i}{dt}.
\end{equation}
Here, $G_i$ denotes either the DNS modal energy gain ($G_E$) or the linear analysis amplification factor ($G_L$), and $\omega$ denotes the RTI growth rate, analogous to the approach of \citet{chang2023}. For the cryostratified regime (figure~\ref{fig:csrun1-compare}d), there is close agreement between $\omega(G_E)$ and $\omega(G_L)$ after the onset of instability growth ($t-t_0\gtrsim7.5$) until RTI formation at $t_1-t_0\approx12$. The agreement is slightly weaker for the cryomictic case, but as $t\rightarrow t_1$, the growth rates still converge until convection begins. 

Further, in figures~\ref{fig:csrun1-compare}e, f, g (cryostratified) and \ref{fig:cmrun1-compare}e, f, g (cryomictic), the normalised modal structures ($\hat{u}$, $\hat{w}$, $\hat{T}$) are compared at the time of RTI formation ($t_1$) for the representative most unstable wavenumber, $\tilde{k}_{\max}(t_1)$. There is excellent agreement between the eigenfunctions of the linear analysis and the Fourier modes of the simulations, notably capturing the difference in structure, and thereby location of RTI formation, between the two regimes due to the difference in unstable wavenumbers (figures~\ref{fig:csrun1-compare}a and \ref{fig:cmrun1-compare}a). Specifically, RTIs are predicted (the depth of peak amplitude of $\hat{w}$) to form in the thermocline at $z\approx-0.5$ for the cryostratified case (figure~\ref{fig:csrun1-compare}g), in agreement with the observed structures in figure~\ref{fig:KE-t1}d. Similarly, for the cryomictic regime, RTIs are predicted to form in the hypolimnion at $z\approx-2$ (figure~\ref{fig:cmrun1-compare}g) which agrees with figure~\ref{fig:KE-t1}b. In summary, for both simulations, the linear analysis accurately captures the early RTI growth dynamics. 

\begin{figure}
    \centering
    \includegraphics[width=\linewidth]{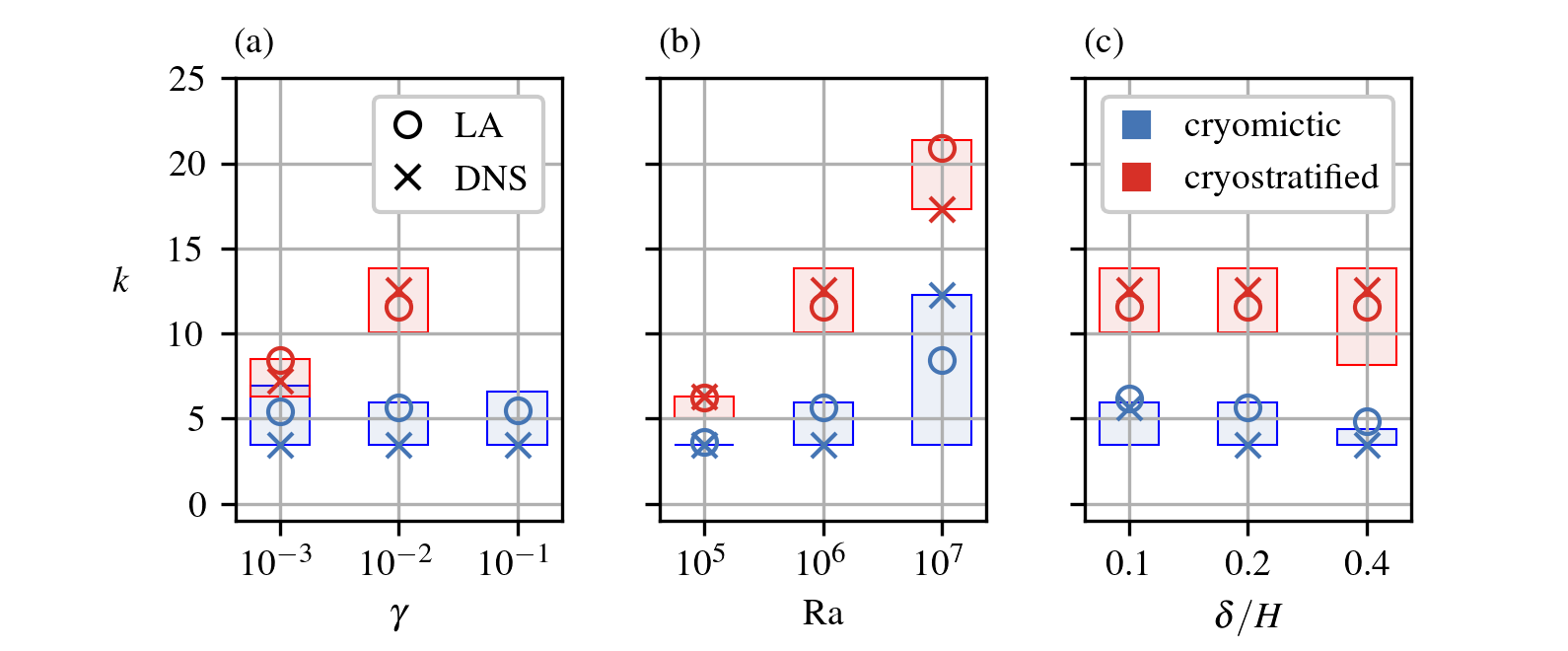}
    \caption{\label{fig:kmax} Comparison of the most unstable wavenumber at $t_1$ as predicted by linear analysis ($\tilde{k}_{\max}$, circles) to DNS ($k_{\max}$, crosses) for (a) varying $\gamma$ with $\Ra=10^6$ and $\delta/H=0.2$, (b) varying $\Ra$ with $\gamma=10^{-2}$ and $\delta/H=0.2$, and (c) varying $\delta/H$ with $\gamma=10^6$ and $\Ra=10^6$. Coloured boxes denote bands of wavenumbers between the smallest and largest $k$ such that $G_E(k,t_1)\geq 0.75\max\{G_E(k,t_1)\}$. Note, there is no $\gamma=10^{-1}$ data since there was no observed $t_1$.}
\end{figure}

To further test the linear analysis, the forcing parameter ($\gamma$), Rayleigh number ($\Ra$) and curvature term ($\delta/H$) are varied. In figure~\ref{fig:kmax}, the representative most unstable wavenumber, $\tilde{k}_{\max}(t_1)$, is compared to the observed most unstable wavenumber, $k_{\max}(t_1)$, at RTI formation. Overall, there is good agreement for all simulations, with the cryostratified configurations mostly presenting a slightly better fit. There is no data point for the cryostratified $\gamma=0.1$ configuration since no RTIs form over the considered timespan ($0\leq t-t_0\leq25$). Both regimes exhibit notable sensitivity to $\Ra$, with better agreement for low $\Ra$ and worse agreement with increasing $\Ra$. As discussed prior, the linear analysis does not typically identify the exact most unstable wavenumber as it is sensitive to the initial noise profile (Appendix~\ref{app:noise}). This uncertainty is estimated by calculating the distance between the smallest and largest $k$ (coloured boxes in figure~\ref{fig:kmax}) such that $G_E(k,t_1)\geq 0.75\max\{G_E(k,t_1)\}$, or in other words, the top $25\%$ gain in modal energy. Note that under this definition, it is not necessarily true that the representative most unstable wavenumber satisfies $G_E(\tilde{k}_{\max},t_1)>0.75\max\{G_E(k,t_1)\}$, but it provides a rough estimate of the span of this unstable band. For all simulations performed here, the linear analysis predicts a wavenumber within this uncertainty band. 

\begin{figure}
    \centering
    \includegraphics[width=\linewidth]{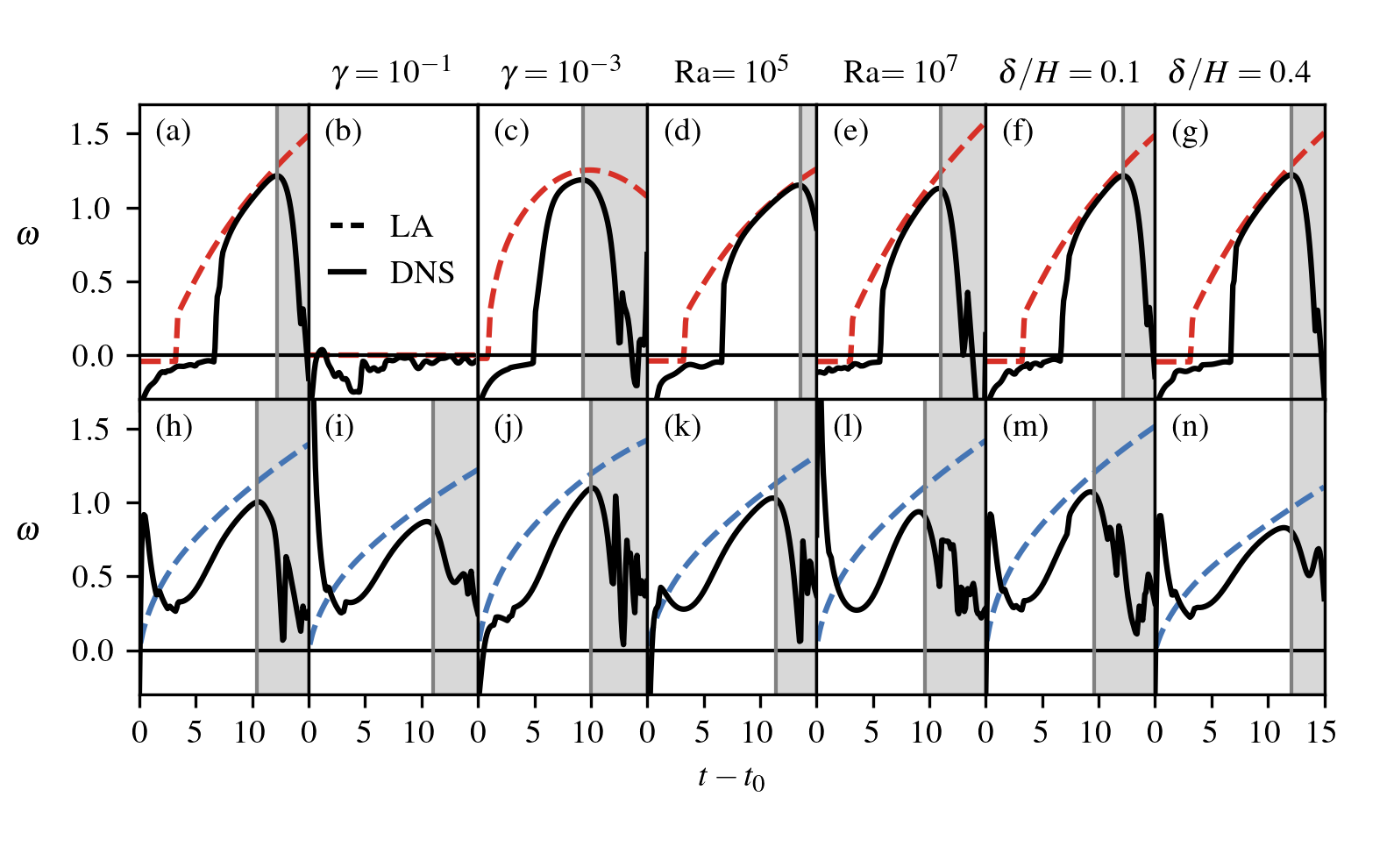}
    \caption{\label{fig:omega} Comparison of the RTI growth rate ($\omega$) of the most unstable wavenumber from the DNS (solid black lines) to the linear analysis (dashed coloured lines) for the cryostratified (top row, red) and cryomictic regimes (bottom row, blue). Each column corresponds to a variance of the listed parameter from the baseline configuration (a), (h) $\gamma=0.01$, $\Ra=10^6$ and $\delta/H=0.2$ (see table~\ref{tab:cases}). Grey translucent boxes denote times after RTI formation ($t\geq t_1$).  }
\end{figure}

In figure~\ref{fig:omega}, the relative RTI growth rate, $\omega$, is compared between the numerical simulations ($k_{\max}$) and linear analysis ($\tilde{k}_{\max}$) for all configurations (table~\ref{tab:cases}). In figures~\ref{fig:omega}a and ~\ref{fig:omega}h, $\omega$ is plotted for the baseline parameters $\gamma=0.01$, $\Ra=10^6$ and $\delta/H=0.2$. For the remaining subplots, the parameter given above in each column is varied for the cryostratified (top row, figures~\ref{fig:omega}a to g) and cryomictic (bottom row, figures~\ref{fig:omega}h to n) regimes. For the cryostratified configurations, once instability sets in after the initial decay period, the linear analysis accurately predicts the growth rate until RTIs form at time $t_1$ (grey lines). There is no observed $t_1$ for the cryostratified $\gamma=0.1$ case and this lack of growth is captured by the linear analysis (figure~\ref{fig:omega}b). For the cryomictic configurations, the agreement is not as strong but still provides a close estimate for $\omega$ around the time of onset ($t_1$). The initial disagreement between the two may be because the linear analysis captures the single most unstable mode, while the numerical simulations feature competing unstable modes (figure~\ref{fig:kmax}) and only the mode with the largest amplification at each time is recorded. As time progresses, the majority of modes dampen and only the most unstable ones amplify (Appendix~\ref{app:noise}), improving agreement between the linear analysis and DNS. The predictions diverge from the simulations after RTI formation at $t=t_1$, consistent with the dynamics becoming increasingly nonlinear. Notably, the observed $t_1$ in the DNS is marked by turning points in the growth rate (following how we define $t_1$ in section \ref{sec:dns-results}). This critical point is generally not captured by the linear analysis. 

\begin{figure}
    \centering
    \includegraphics[width=\linewidth]{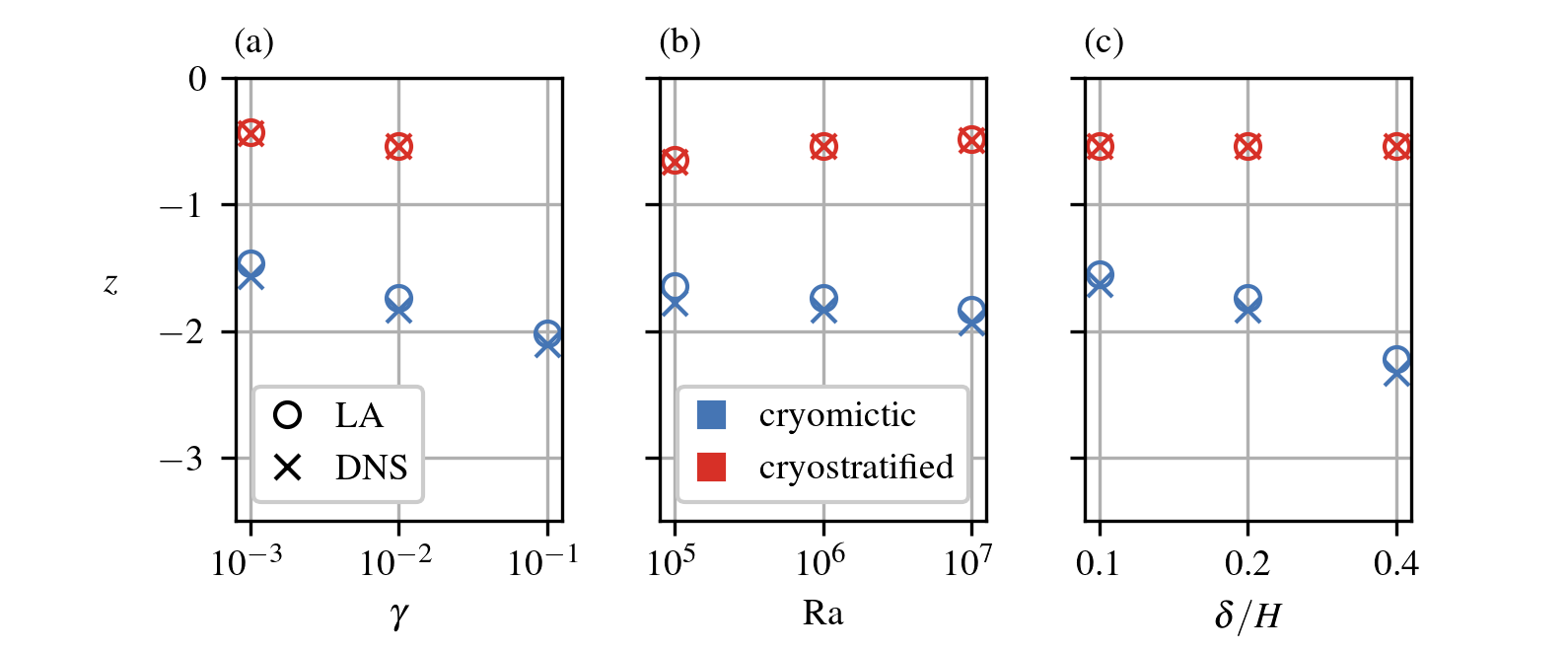}
    \caption{\label{fig:z0} Comparison of the depth at which RTIs form at time $t_1$ between the simulations (DNS, crosses) and predictions (linear analysis, circles) for the cryostratified (red) and cryomictic (blue) regimes.}
\end{figure}

The depth at which RTIs are observed in the numerical simulations are compared to predictions made by the linear analysis in figure~\ref{fig:z0}. For the simulations, this depth is estimated from the vertical velocity field at time $t_1$ (e.g., figures~\ref{fig:KE-t1}b and d), computing the horizontal mean of $|\tilde{w}|$. The depth at which this reaches its maximum is interpreted as the depth at which RTIs form in the DNS. For the predictions, we take the depth at which the vertical velocity modal amplitude, $\hat{w}$, peaks for the most unstable mode $\tilde{k}_{\max}(t_1)$ (e.g., figures~\ref{fig:csrun1-compare}f and figures~\ref{fig:cmrun1-compare}f). Figure~\ref{fig:z0} shows consistently close agreement between the simulations and predictions.  

Comparing the cryostratified and cryomictic regimes in the figures above, clear differences can be seen. First, the cryostratified configurations have larger unstable wavenumbers (figure~\ref{fig:kmax}) indicating that smaller wavelengths can grow to form a greater number of smaller scale RTIs (figures~\ref{fig:DNS} and \ref{fig:KE-t1}) potentially leading to a more intense initial mixing layer. This is supported by the accelerated modal growth (figure~\ref{fig:omega}) for the cryostratified cases which reach larger maxima ($\omega>1$) compared to the cryomictic regime ($\omega\lesssim1$). Although RTIs form earlier for the cryomictic cases (figure~\ref{fig:t1}), in all the cryostratified cases (not including $\gamma=0.1$, figure~\ref{fig:omega}b) a period of decay is observed until instabilities can overcome diffusive processes within the thermocline and begin to grow. Hence, perturbations in the cryomictic regime grow instantly without a decay period resulting in more initial energy once instabilities begin to grow and thereby RTIs form quicker. The variance in wavenumber between the two regimes also affects the dominant modal structure (figures~\ref{fig:csrun1-compare}f and \ref{fig:cmrun1-compare}f) and thereby the depth at which RTI formation occurs (figure~\ref{fig:z0}). For instance, RTIs always form within the thermocline ($z>-H/\lambda=-5$) for the cryostratified regime because light absorption is largely confined there. For the cryomictic regime, RTIs form within the hypolimnion ($z<-H/\lambda=-1$) as the shallower thermocline allows light to penetrate into the hypolimnion. 

When varying the forcing ($\gamma$), the unstable modes (figure~\ref{fig:kmax}a) in the cryomictic regime are uninfluenced whereas in the cryostratified regime, the most amplified wavenumber increases with $\gamma$ (less forcing). One reason for this difference in behaviour could be due to an increase of advective heat transport in the thermocline, where RTIs form in the cryostratified regime (figure~\ref{fig:z0}), as $\gamma$ is increased (see equation~\ref{eq:lin_NS_heat}). This may transport large wavelength (small $k$) perturbations allowing smaller wavelengths (large $k$) to grow. Whereas, for the cryomictic configurations, instabilities in the hypolimnion are unaffected due to the isothermal approximation ($\partial\bar{T}/\partial z\approx0$). In both regimes, we generally observe that decreasing $\gamma$ increases the growth rate due to the higher level of solar forcing (figures~\ref{fig:omega}c and j), notably shortening the decay period in the cryostratified regime (figure~\ref{fig:omega}c). Increasing $\gamma$ decreases the growth rate due to the reduction in forcing (figures~\ref{fig:omega}i) and thickens the thermal boundary layer \citep{Smith2025} leading to deeper RTIs (figure~\ref{fig:z0}a). This latter effect is more amplified for the cryomictic regime as there are no thermal gradients to overcome in the hypolimnion and so increased thermal diffusion allows heat to travel deeper. 

Increasing $\Ra$ leads to larger unstable wavenumbers (figure~\ref{fig:kmax}b). This is due to the decrease in dissipation for increased $\Ra$, allowing finer scale structures to emerge with shorter wavelengths, thereby expanding the range of unstable wavenumbers. Further, this decrease in dissipation leads to steeper, or more accelerated, growth rates (figures~\ref{fig:omega}d, e, k and l) and slightly shallower RTIs for the cryostratified configurations (figure~\ref{fig:z0}b) likely due to a thinning of the boundary layer. This latter point is not observed for the cryomictic regime suggesting that RTIs form too deep to interact with any boundary layers. 

Lastly, varying the curvature ($\delta/H$) has little-to-no impact on the cryostratified configurations as RTIs form within the thermocline (figure~\ref{fig:z0}) away from curvature effects. For the cryomictic regime, where RTIs form within the hypolimnion, a more extended transition region (i.e. higher $\delta/H$) results in a decrease in growth rate (figures~\ref{fig:omega}m and n) and deeper RTIs (figure~\ref{fig:z0}c) due to the amplification of thermal gradients. 

\subsection{Predicting Onset of RTI Formation ($t_1$)}

The previous subsection has shown that the linear analysis approximates the early RTI growth regime (figures~\ref{fig:kmax}--\ref{fig:z0}) fairly well. As a result, we apply our linear analysis here to predict the time when RTIs form and initiate convection ($t_1$). 

The onset of RTI formation is estimated by considering the amplification $G_L$ \eqref{eq:Gtk} of the representative most unstable mode at time $t^\dagger \equiv t_0+15$, i.e. $\tilde{k}_{\max}(t^\dagger)$. This time is chosen since the most unstable wavenumber does not vary significantly (figures~\ref{fig:csrun1-compare}c and \ref{fig:cmrun1-compare}c) and so we expect little variation between the growth history compared to other possibly more unstable modes. Selecting this approximate most unstable mode is intended to balance accuracy with computational cost. Further, we note that in our simulations $G_L \approx 10^3$ at $t_1$ (figures~\ref{fig:csrun1-compare} and \ref{fig:cmrun1-compare}) ranging approximately between $4.5\times10^2$ and $3.5\times10^3$. Consequently, $t_1$ is approximated by the minimum amount of time for the most unstable mode at $t^\dagger$ to amplify by $10^3$, or
\begin{equation}\label{eq:t1_threshold}
    G_L\left(\tilde{k}_{\max}(t^\dagger), t_1\right) = 10^3.
\end{equation}
This threshold is used as a practical reference level rather than a precise criterion for the onset of RTIs since nonlinear effects are expected to become non-negligible at $t_1$ (figures~\ref{fig:KE-t1} and \ref{fig:omega}), and the linear approximation becomes less physically meaningful. Although the threshold is slightly dependent on the scale of the initial perturbations (Appendix~\ref{app:noise}), the predictions still remain fairly robust. Comparable thresholds have been applied before in studies of RDC in ice-free lakes \citep[e.g.][]{yang2026}. 

\begin{figure}
    \centering
    \includegraphics[width=\linewidth]{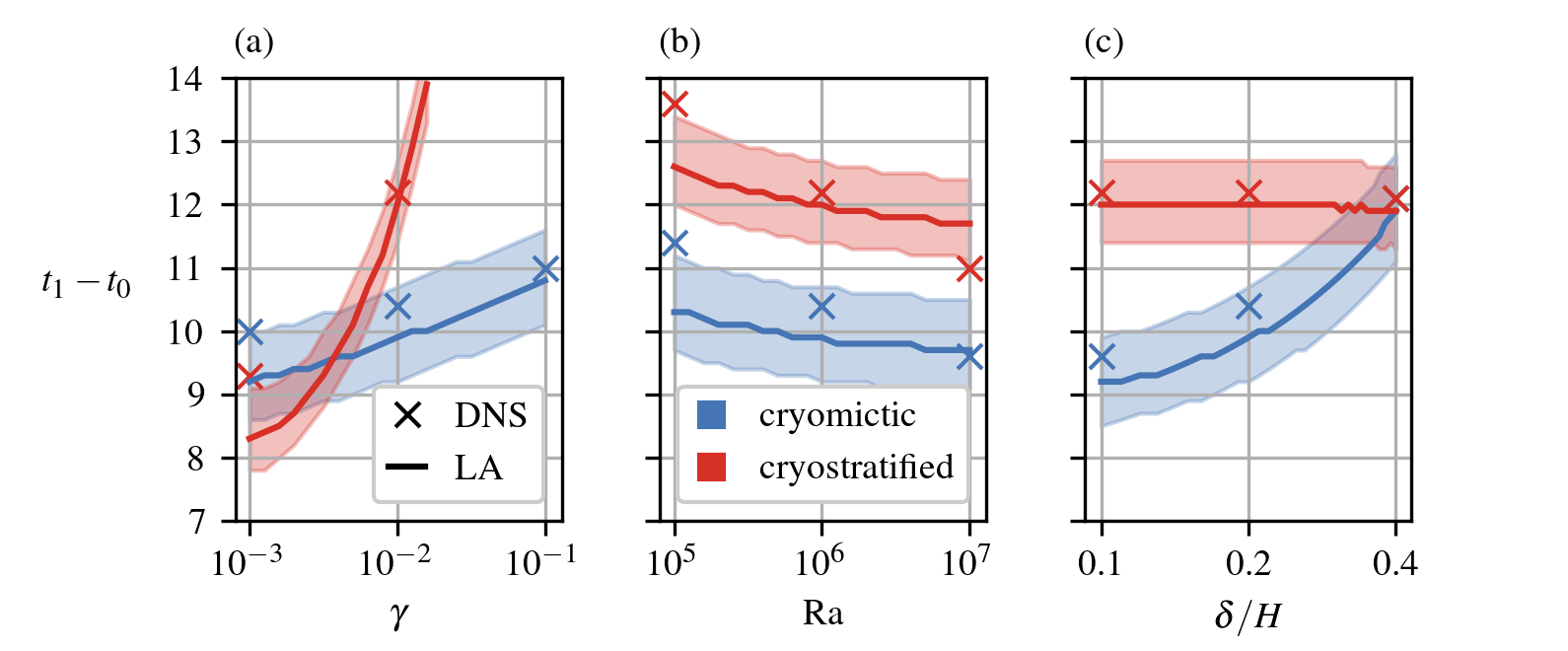}
    \caption{\label{fig:predict_t1} Comparison of RDC onset ($t_1$) relative to gravitational instability ($t_0$) between the numerical simulations (crosses) and predictions by linear analysis (lines) for the cryostratified ($\lambda/H=0.2$, $T_H/\TMD=1$, red) and cryomictic ($\lambda/H=1$, $T_H/\TMD=0.2$, blue) regimes. Predictions for $t_1$ are made once the amplification of the representative most unstable wavenumber at $t-t_0=15$ is greater than 1000, i.e. $G(\tilde{k}_{\max}(t_0+15),t_1)\geq1000$. Translucent coloured areas surrounding the predictions correspond to the range of $t_1$ between the thresholds $500$ (lower curve) and $2500$ (upper curve).}
\end{figure}

In figure~\ref{fig:predict_t1}, our predictions for RDC onset are compared to the DNS results from figure~\ref{fig:t1} for both the cryomictic and cryostratified regimes where there is good overall agreement. Uncertainty is introduced by the threshold applied to $G_L(t_1)$ as discussed above, as well as the inherent entropy within the system (due to, for example, the random initial noise distribution). An estimate of uncertainty in $t_1$ is provided by considering a range of thresholds, $G_L = [500,2500]$ in figure~\ref{fig:predict_t1}. The DNS data points generally tend to lie within this range suggesting that $1000$ is an appropriate threshold for $G_L(t_1)$. 

As discussed before, increasing $\gamma$ (figure~\ref{fig:predict_t1}a) increases the time to RTI formation as forcing is reduced and dampens the growth rate (figures~\ref{fig:omega}b and i). Conversely, decreasing $\gamma$ increases the growth rate (figures~\ref{fig:omega}c and j) due to the gain in forcing. Notably, the transition point is observed at $\gamma\approx4\times10^{-3}$ where RTIs in cryostratified lakes grow quicker and form before cryomictic lakes (relative to $t_0$). By increasing $\Ra$ (figure~\ref{fig:predict_t1}b) there is a decrease in the time until RTI formation. This is because of the higher rate of growth (figures~\ref{fig:omega}d and k) that results from the decrease in dissipation. Lastly, as before, varying curvature ($\delta/H$, figure~\ref{fig:predict_t1}c) has little effect on the cryostratified regime since RTIs are confined to the thermocline (figure~\ref{fig:z0}). Increasing curvature has a bigger effect for the cryomictic regime where thermal gradients are amplified and damps growth (figures~\ref{fig:omega}m and n).   

Recall that for all configurations that go unstable, $t^*_1-t^*_0\sim O(10\tau)$ (figure~\ref{fig:predict_t1}) suggesting for cryomictic lakes $t^*_1\approx10\tau$. For cryostratified lakes, we can apply the small-$\gamma$ approximation $t^*_0 \approx e^{5.3\sqrt{\gamma}}$ from \cite{Smith2025}, which was found to perform well here (table~\ref{tab:t0}). We therefore propose the following cryostratified approximation:
\begin{equation}\label{eq:t1_estimate}
    t^*_1 \approx \left(e^{5.3\sqrt{\gamma}} + \frac{10}{\gamma\sqrt{\Ra\Pr}}\right)\tau_f.
\end{equation}
This reduces to the estimate $t^*_1/\tau_f = e^{5.3\sqrt{\gamma}} +1$ proposed by \citet{Smith2025} for $\Ra=10^5$ and $\gamma=0.01$. Further, it allows an estimation of $t^*_1$ without conducting the full linear analysis.

\subsection{Application to Field Measurements}

\begin{figure}
    \centering
    \includegraphics[width=\linewidth]{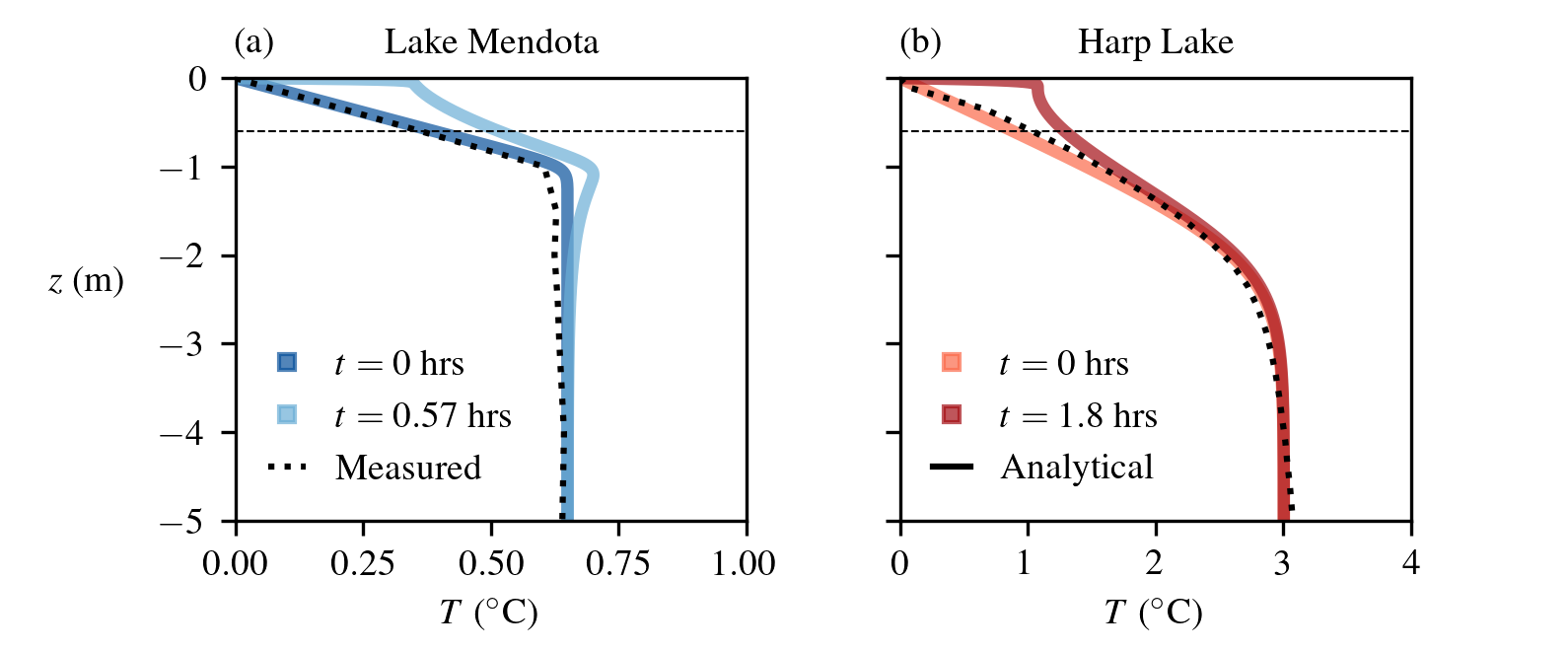}
    \caption{\label{fig:laket0} Estimates of temperature profiles (solid lines) against the measured profiles (dotted black lines) from figure~\ref{fig:lake-profiles} for (a) Lake Mendota, WI, USA and (b) Harp Lake, ON, CA. Estimates use the analytical solution \eqref{eq:1DforcingT0sol} assuming moderate-to-clear water ($\lambda=0.6$ m) and high forcing ($Q_0=10^{-4}\celsius$ m s$^{-1}$) with (a) $T_H=0.65\celsius$, $H=1$ m, $\delta=0.1$ m and (b) $T_H=3\celsius$, $H=2$ m, $\delta=0.8$ m.}
\end{figure}

Here, we apply our framework to predict the onset of RDC for cryomictic Lake Mendota and cryostratified Harp Lake. Using equation \eqref{eq:DICs}, the initial profiles in figure~\ref{fig:lake-profiles} are approximated by $H=1$ m, $T_H=0.65\celsius$ and $\delta=0.1$ m for Lake Mendota (figure~\ref{fig:laket0}a), and $H=2$ m, $T_H=3\celsius$ and $\delta=0.8$ m for Harp Lake (figure~\ref{fig:laket0}b). For both lakes, moderate-to-clear water clarity ($\lambda=0.6$ m) is assumed with typical cold water properties: $\nu\approx10^{-6}$ m$^2$ s$^{-1}$, $\kappa\approx10^{-7}$ m$^2$ s$^{-1}$ and $c_p\approx4210$ J kg$^{-1}\celsius^{-1}$. 

To estimate $t_1^*$, the same process outlined in the previous subsection is applied including the use of the analytical solution \eqref{eq:1DforcingT0sol} to calculate $t_0^*$. For moderate forcing ($Q_0=10^{-5}\celsius$ m s$^{-1}$), such as for typical late-winter ice covers with a mixture of black and white ice, the model predicts that the gravitationally unstable layer emerges within $t_0^*\approx28$ hours for Harp Lake with convection 90 minutes later to give $t_1^*\approx30$ hours. For Lake Mendota, convection is predicted after $t_1^*\approx80$ minutes. For higher forcing ($Q_0=10^{-4}\celsius$ m s$^{-1}$) such as on sunny days or for thin transparent ice, Harp Lake becomes unstable after $t_0^*\approx1.8$ hours and convection begins within 26 minutes to give $t_1^*\approx2.3$ hours (figure~\ref{fig:laket0}b), whilst for Lake Mendota, convection begins after $t_1^*\approx34$ minutes (figure~\ref{fig:laket0}a). For low forcing ($Q_0=10^{-6}\celsius$ m s$^{-1}$) on darker days or ice covers made of opaque white ice, Harp Lake only becomes unstable after $t_0^*\approx1390$ hours or 58 days but the model predicts a substantial decay period over several hours and convection is not predicted in the DNS due to insufficient forcing ($\gamma\approx0.1$). However for cryomictic Lake Mendota, the forcing is sufficient and the model predicts convection after $t_1^*\approx3$ hours. 

The estimates above compare well with equation~\eqref{eq:t1_estimate}, for example with moderate forcing ($Q_0=10^{-5}\celsius$ m s$^{-1}$), $t^*_1\approx30$ hours for Harp Lake and $t^*_1\approx70$ minutes for Lake Mendota. This improves as the forcing is increased to $Q_0=10^{-4}\celsius$ m s$^{-1}$ where $t_1^*\approx2.4$ hours and $t_1^*\approx30$ minutes for Harp Lake and Lake Mendota respectively. However, as the forcing is decreased to $Q_0=10^{-6}\celsius$ m s$^{-1}$, agreement slightly decreases, giving $t_1^*\approx2.4$ hours for Lake Mendota. For Harp Lake, the small $\gamma$ assumption is violated and so we do not expect good agreement, but equation~\eqref{eq:t1_estimate} estimates $t_1^*\approx1171$ hours which is significantly shorter than the calculated $t_0^*\approx1390$ hours.  

In summary, equation~\eqref{eq:t1_estimate} provides a reasonable estimate for $t_1^*$ as long as there is sufficient forcing (small $\gamma$). It is likely that Lake Mendota would begin convection and mixing within a single day whereas Harp Lake would require several days to develop the unstable layer, but afterwards convection would likely only take a few hours. However this does not consider the diurnal cycle which would impact these estimates.  

%%%%%%%%%%%%%%%%%%%%%%%%%%%%%%%%%%%%%%%%%%%%%%%%%%%%%%%%%%%%%%%%
%%%%%%%%%%%%%%%%%%%%%%%%%%%%%%%%%%%%%%%%%%%%%%%%%%%%%%%%%%%%%%%%

\section{Conclusions}\label{sec:conc}

Overall, the results demonstrate the success of the linear analysis in predicting the timing, location and structure of RTIs, and allows for further understanding of environmental influences. First, there is generally distinct behaviour between the two types of thermal structures, in particular, cryostratified lakes will have a long period of quiescence ($t^*_0\gg\tau$, table~\ref{tab:t0}) where a gravitationally unstable layer must form in the thermocline before RTIs can begin to grow (figure~\ref{fig:analytic-sols}). This is a natural consequence of the narrow concentration of light absorption in the thermocline with negligible interaction with the deeper isothermal hypolimnion. In contrast, the cryomictic case has a broader spread of light absorption, and significant heating occurs in the hypolimnion. This leads to quick emergence of gravitational instability ($t_0\sim0$) and allows for rapid RTI growth (figure~\ref{fig:DNS}). As a result, RTI formation at time $t_1$ will take substantially longer for cryostratified lakes. Even if this period of quiescence is taken into account and we consider dynamics when both lake types are unstable ($t>t_0$), generally, cryostratified lakes will still take longer until RTIs form (figure~\ref{fig:t1}). This is due to another, albeit much shorter, period of decay for the cryostratified regime (figure~\ref{fig:omega}). This period is a result of the stabilising properties of the thermocline, which also determines which wavelengths can grow (figure~\ref{fig:kmax}). The cryostratified regime always has comparatively smaller unstable wavelengths to the cryomictic regime, which in turn influences the quantity (figures~\ref{fig:DNS} and \ref{fig:KE-t1}) and depth (figure~\ref{fig:z0}) of RTIs. 

Further, the solar forcing ($Q_0$) was adjusted by varying $\gamma$. Increasing the forcing (decreasing $\gamma$) led to higher growth rates (figure~\ref{fig:omega}), shorter $t_1$ (figure~\ref{fig:t1}a) and shallower RTIs (figure~\ref{fig:z0}a). Conversely, decreasing the forcing (increasing $\gamma$) led to lower growth rates, increased $t_1$ and deeper RTIs. Although $\Ra$ is a function of the background temperature ($\Delta T$), solar forcing ($Q_0$) and water clarity ($\lambda$), we interpret changes in $\Ra$ predominantly as a change in water clarity due to the dominance of the 8/3 exponent in equation~\eqref{eq:Ra}. However, this is not equivalent to simply increasing $\lambda$ since $\lambda/H$ (the location/concentration of heating) is fixed, and so by increasing $\Ra$, the thermocline depth ($H$) is simultaneously increased. Hence, interpreting changes in $\Ra$ as changes in water clarity needs careful consideration. Increasing $\Ra$ resulted in smaller unstable wavelengths (figure~\ref{fig:kmax}b) due to the decreased rates of viscous and thermal dissipation which allows higher rates of growth (figure~\ref{fig:omega}) and thereby shorter $t_1$ (figure~\ref{fig:t1}b). On the other hand, by decreasing $\Ra$ there are larger unstable wavelengths, decreased rates of growth and longer $t_1$ due to the higher rates of dissipation. 

In summary, we have developed an idealised framework to investigate the onset of ice-covered RDC in cryomictic and cryostratified lakes \citep{yang2021}. Two-dimensional DNS and linear analysis has allowed the exploration of how environmental factors such as stratification, water clarity and solar forcing impact the the time-scales and growth rates associated with the formation of RTIs. By applying linear analysis, a model of the early RTI growth phase was developed with a temporally evolving background whose predictions for the onset of RTI formation compare well with simulations. More specifically, once the water column is sufficiently gravitationally unstable ($t>t_0$), the model accurately captures which wavelengths grow (figure~\ref{fig:kmax}), their associated growth rates (figure~\ref{fig:omega}) and at which depth RTIs form (figure~\ref{fig:z0}). As a result, a threshold \eqref{eq:t1_threshold} was proposed based on the amplification of the finite-time propagator \eqref{eq:finite_time_propagator} that allows estimations of $t_1$ which compares well to simulations (figure~\ref{fig:predict_t1}). Although the linear model can accurately characterise the early RTI growth dynamics, it is difficult under this framework to exactly pinpoint when RTIs form. For instance, $t_1$ is measured from the simulations by taking the maximum in the kinetic energy growth rate (figure~\ref{fig:KE-t1}) which results from nonlinear dynamics (e.g. rotation and shear). Hence, in future, these effects need to be incorporated to make more accurate predictions, however, current results already provide a reasonable approximation (figure~\ref{fig:predict_t1}).     

Going forward, including other environmental processes such as the diurnal cycle \citep[e.g.][]{chang2023} will be needed to more accurately constrain the onset of RDC in realistic settings. Further, additional numerical simulations, high-frequency field observations including measurements of small scale noise (e.g. wind-driven oscillations), and lab-scale experiments are needed to verify our analysis. Lastly, incorporating melting would provide interesting details about how melt rates are coupled to RDC and how both processes interact to determine lake ice break-up.

%%%%%%%%%%%%%%%%%%%%%%%%%%%%%%%%%%%%%%%%%%%%%%%%%%%%%%%%%%%%%%%%
%%%%%%%%%%%%%%%%%%%%%%%%%%%%%%%%%%%%%%%%%%%%%%%%%%%%%%%%%%%%%%%%

\backsection[Supplementary data]{\label{SupMat}Supplementary material and movies are available at [to be filled upon acceptance]}

\backsection[Acknowledgements]{The authors thank Samuel Salemink-Harry, George Brooks, Hilary Dugan, Martina Rogers, Grace Wilkinson and Lucas Zoet for insightful discussions. The authors wish to recognise the \citet{HPC} at the University of Wisconsin-Madison for the computational resources to support this study.}

\backsection[Funding]{Support for this research was provided by the University of Wisconsin–Madison Office of the Vice Chancellor for Research with funding from the Wisconsin Alumni Research Foundation.}

\backsection[Declaration of interests]{The authors report no conflict of interest.}

\backsection[Data availability statement]{The data that support the findings of this study are openly available in [to be filled upon acceptance]
%[repository name] at http://doi.org/[doi], reference number [reference number]. See JFM's \href{https://www.cambridge.org/core/journals/journal-of-fluid-mechanics/information/journal-policies/research-transparency}{research transparency policy} for more information
}

\backsection[Author ORCIDs]{B. J. Smith, https://orcid.org/0009-0000-3695-1471; B. Yan, https://orcid.org/0000-0002-4036-8577; T. J. W. Wagner, https://orcid.org/0000-0003-4572-1285; N. Pujara, https://orcid.org/0000-0002-0274-4527; J. A. Franck, https://orcid.org/0000-0001-8456-5153}

\backsection[Author contributions]{J.F., N.P., B.S., B.Y., and T.J.W.W. designed the study, B.S. conducted the numerical simulations, B.Y. derived the linear analysis, B.S. and B.Y. performed data analysis, all authors contributed to the interpretation of results and the writing of the manuscript.}

%%%%%%%%%%%%%%%%%%%%%%%%%%%%%%%%%%%%%%%%%%%%%%%%%%%%%%%%%%%%%%%%
%%%%%%%%%%%%%%%%%%%%%%%%%%%%%%%%%%%%%%%%%%%%%%%%%%%%%%%%%%%%%%%%
\appendix
%%%%%%%%%%%%%%%%%%%%%%%%%%%%%%%%%%%%%%%%%%%%%%%%%%%%%%%%%%%%%%%%
%%%%%%%%%%%%%%%%%%%%%%%%%%%%%%%%%%%%%%%%%%%%%%%%%%%%%%%%%%%%%%%%
\section{Linearised Radiatively Driven Convection}\label{app:LA}
%%%%%%%%%%%%%%%%%%%%%%%%%%%%%%%%%%%%%%%%%%%%%%%%%%%%%%%%%%%%%%%%

\subsection{Linear Perturbation Equations}\label{app:LA-pert-eqns}

During the initial linear growth stage, the magnitude of transient perturbations are generally much smaller compared to the characteristic scales of the background. Hence, it is appropriate to introduce a separate set of scales for the perturbation quantities. The flow variables are first decomposed into background quantities (bars) and small perturbations (tildes):
\begin{equation}\label{eq:lin_decomp}
\vec{u} = \vec{0} + \tilde{\vec{u}}, \quad P = \bar{P}(z,t) + \tilde{P}, \quad T = \bar{T}(z,t) + \tilde{T}.
\end{equation}
The water is assumed to be initially quiescent ($\bar{\vec{u}}\equiv0$) where the background satisfies
\begin{subequations}\label{eq:background}
    \begin{gather}\label{eq:1DheatD}
        \frac{\partial \bar{T}}{\partial t} = \kappa\frac{\partial^2\bar{T}}{\partial z^2} + \frac{Q_0}{\lambda}e^\frac{z}{\lambda}, \\[5pt]
        \frac{\partial \bar{P}}{\partial z} = \rhoMD Cg(\bar{T}-\TMD)^2.
    \end{gather}
\end{subequations}
Substituting these into \eqref{eq:NSD}, and neglecting higher-order perturbations yields the linearised system,
\begin{subequations}\label{eq:lin_NSD}
    \begin{gather}
        \bnabla \bcdot \tilde{\vec{u}} = 0, \\[5pt]
        \frac{\partial \tilde{\vec u}}{\partial t}
        = -\frac{1}{\rhoMD}\bnabla \tilde{P} 
        + \nu\nabla^2\tilde{\vec{u}} 
        + 2Cg(\bar{T}-\TMD)\tilde{T}\vec{\hat{e}}_z, \\[5pt]
        \label{eq:lin_heat}
        \frac{\partial \tilde{T}}{\partial t}
        + \tilde{w}\frac{\partial\bar{T}}{\partial z}
        = \kappa\frac{\partial^2\tilde{T}}{\partial z^2},
    \end{gather}
\end{subequations}
where $\tilde{w}$ is the vertical velocity of the perturbations. 

\subsection{Scales in the Linear Regime}\label{app:LA-scales}

The characteristic timescale $\tau$ is defined as the time required for a disturbance to travel a distance $\lambda$ with velocity $W$, or $\tau = \lambda/W$. To determine $W$, we follow \citet{chang2023} and assume that the initial vertical acceleration is governed by buoyancy arising from temperature-induced density variations. By balancing the vertical acceleration with the buoyancy, 
\begin{equation}\label{eq:mom_sim}
    \frac{W}{\tau} \sim 2C g (\TMD - \bar{T})\theta.
\end{equation}
where $\theta$ is the scale of the temperature perturbations ($\tilde{T}$). 

To determine $\theta$, we assume that the temperature gradient responsible for triggering the instability is generated solely by radiation. In the absence of diffusion, then for $t\geq t_0$,
\begin{equation}
    \frac{\partial \bar{T}}{\partial z} = \frac{Q_0}{\lambda^2}(t-t_0)e^{z/\lambda} + \frac{d}{dz}\bar{T}(t=t_0).
\end{equation}
For the onset of RTIs, the relevant requirement is the development of a positive vertical temperature gradient ($\partial\bar{T}/\partial z>0$), meaning that for regions of instability $d\bar{T}(t=t_0)/dz \approx 0$. Therefore, over a characteristic time interval $t-t_0\sim\tau$, the radiation-induced increase in the background temperature gradient is approximately $\partial\bar{T}/\partial z\sim Q_0\tau/\lambda^2$. Further, assuming that the change in temperature is balanced by convective heat transfer in the linearised heat equation \eqref{eq:lin_heat}, 
\begin{equation}
    \frac{\theta}{\tau} \sim \tilde{w}\frac{\partial\bar{T}}{\partial z} \sim W \frac{Q_0}{\lambda^2}\tau = \frac{Q_0}{\lambda},
\end{equation}
and $\theta=Q_0/W$ which is simply the heat gained over a time $\tau$. 

Finally, the velocity scale is obtained by combining both temperature scales in equation \eqref{eq:mom_sim}, recalling that $\bar{T}\sim T_\lambda$, so that
\begin{equation}
W = \proot{2\lambda CgQ_0\Delta T}{3}, \quad \tau = \proot{\frac{\lambda^2}{2CgQ_0\Delta T}}{3}, \quad \theta = \proot{\frac{Q_0^2}{2\lambda Cg\Delta T}}{3},
\end{equation}
where $\Delta T=\TMD-T_\lambda$. These match with the (non-background flow) scalings proposed by \citet{Grace2022} and \citet{chang2023} assuming that the buoyancy flux scales as $B_0\sim\alpha gQ_0$ where the thermal expansion coefficient $\alpha\approx-\rhoMD^{-1}\,d\rho/dT$ scales as $2C\Delta T$. 

%%%%%%%%%%%%%%%%%%%%%%%%%%%%%%%%%%%%%%%%%%%%%%%%%%%%%%%%%%%%%%%%
%%%%%%%%%%%%%%%%%%%%%%%%%%%%%%%%%%%%%%%%%%%%%%%%%%%%%%%%%%%%%%%%
\section{Grid Resolution \& Noise Sensitivity Studies}
%%%%%%%%%%%%%%%%%%%%%%%%%%%%%%%%%%%%%%%%%%%%%%%%%%%%%%%%%%%%%%%%
\subsection{Grid Resolution}\label{app:grid-res}

\begin{figure}
    \centering
    \includegraphics[width=\linewidth]{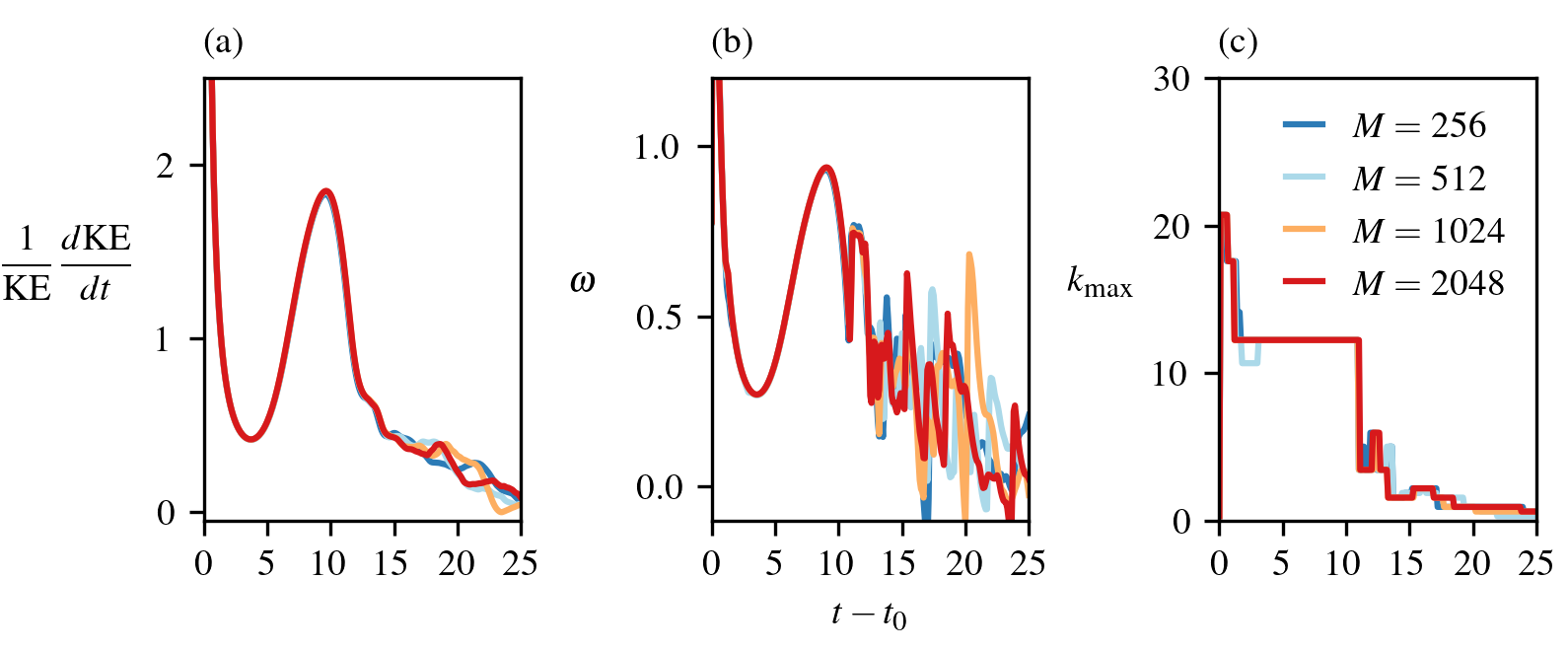}
    \caption{\label{fig:grid-res} Sensitivity of results to the grid resolution ($M$) used in the numerical simulations for the cryomictic ($\lambda/H=1$, $T_H/\TMD=0.2$) regime with $\gamma=10^{-2}$, $\Ra=10^7$ and $\delta/H=0.2$. (a) The kinetic energy (KE, equation~\ref{eq:KEgrowth}) growth rate, (b) modal energy growth rate ($\omega$, equation~\ref{eq:omega}) and (c) most unstable wavenumber ($k_{\max}$, equation~\ref{eq:Gtk}).}
\end{figure}

In figure~\ref{fig:grid-res}, the grid resolution ($M$) is varied in the numerical simulations to investigate the sensitivity of results to the number of points. This is performed on the large Rayleigh number ($\Ra=10^7$) cryomictic case, since RTIs can begin to grow immediately and the expected finer scale dynamics associated with the reduced rates of dissipation. The onset of RTI formation ($t_1$) is estimated once the turning point occurs in the relative kinetic energy growth rate \eqref{eq:KEgrowth}. As seen in figure~\ref{fig:grid-res}a, $t_1$ is insensitive to the number of points. Additionally, results before $t\leq t_1$ are insensitive to $M$ for both the growth rate of modal energy, $\omega$ (figure~\ref{fig:grid-res}b), and the associated most amplified wavenumber, $k_{\max}$ (figure~\ref{fig:grid-res}c). 

\subsection{Noise Sensitivity}\label{app:noise}

\begin{figure}
    \centering
    \includegraphics[width=\linewidth]{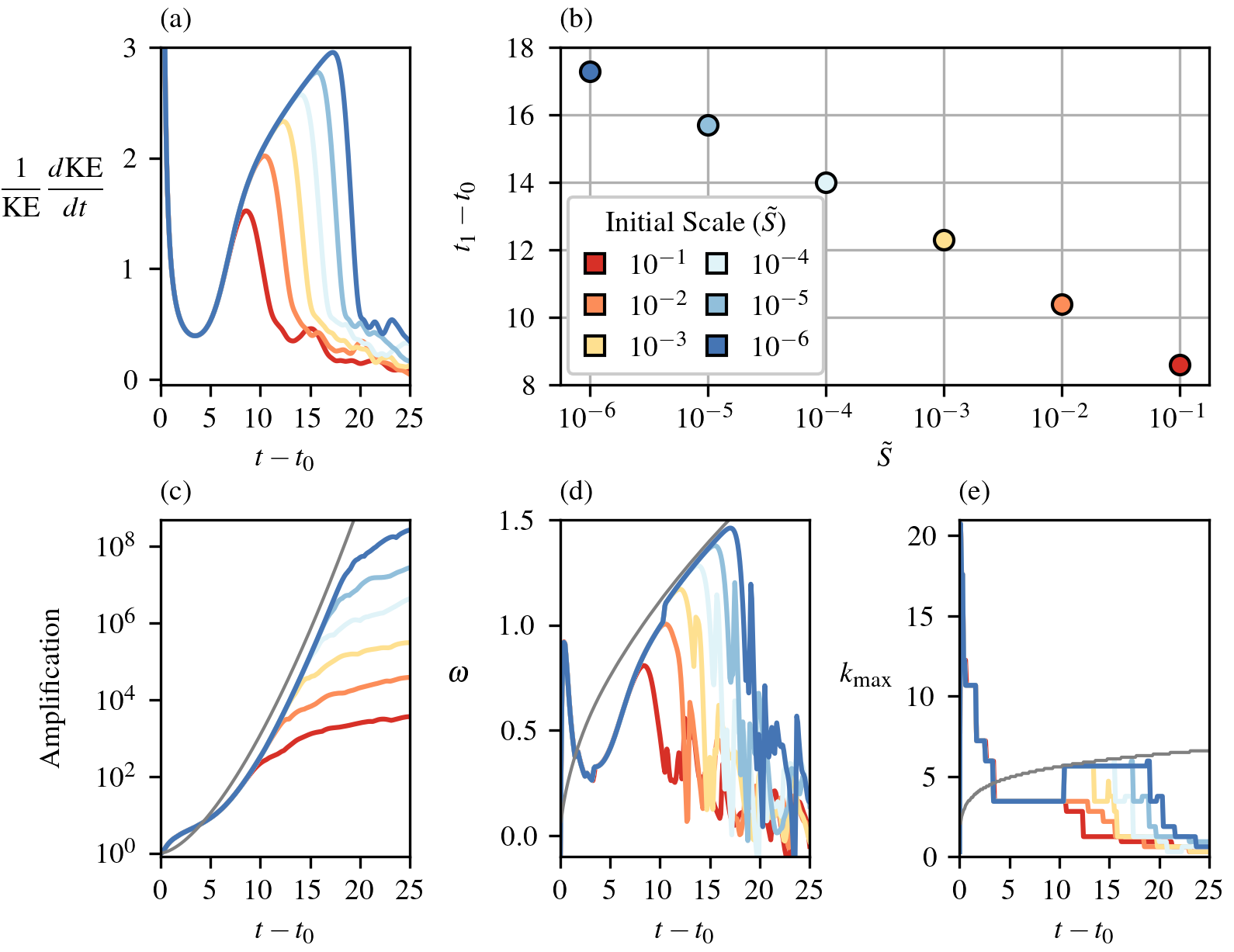}
    \caption{\label{fig:noise-scale} Sensitivity of results to the initial scale ($\tilde{S}$) of perturbations used in the numerical simulations for the cryomictic ($\lambda/H=1$, $T_H/\TMD=0.2$) regime with $\gamma=10^{-2}$, $\Ra=10^6$ and $\delta/H=0.2$. (a) The kinetic energy (KE, equation~\ref{eq:KEgrowth}) growth rate, (b) values of $t_1$ shifted by $t_0$ for each scale of noise, (c) the amplification of the most unstable mode with the prediction by linear analysis in grey, (d) growth rate ($\omega$, equation~\ref{eq:omega}), and (e) most unstable wavenumber ($k_{\max}$, equation~\ref{eq:Gtk}) with the predicted most unstable wavenumber, $\tilde{k}_{\max}(t)$, overlaid in grey.}
\end{figure}

\begin{figure}
    \centering
    \includegraphics[width=\linewidth]{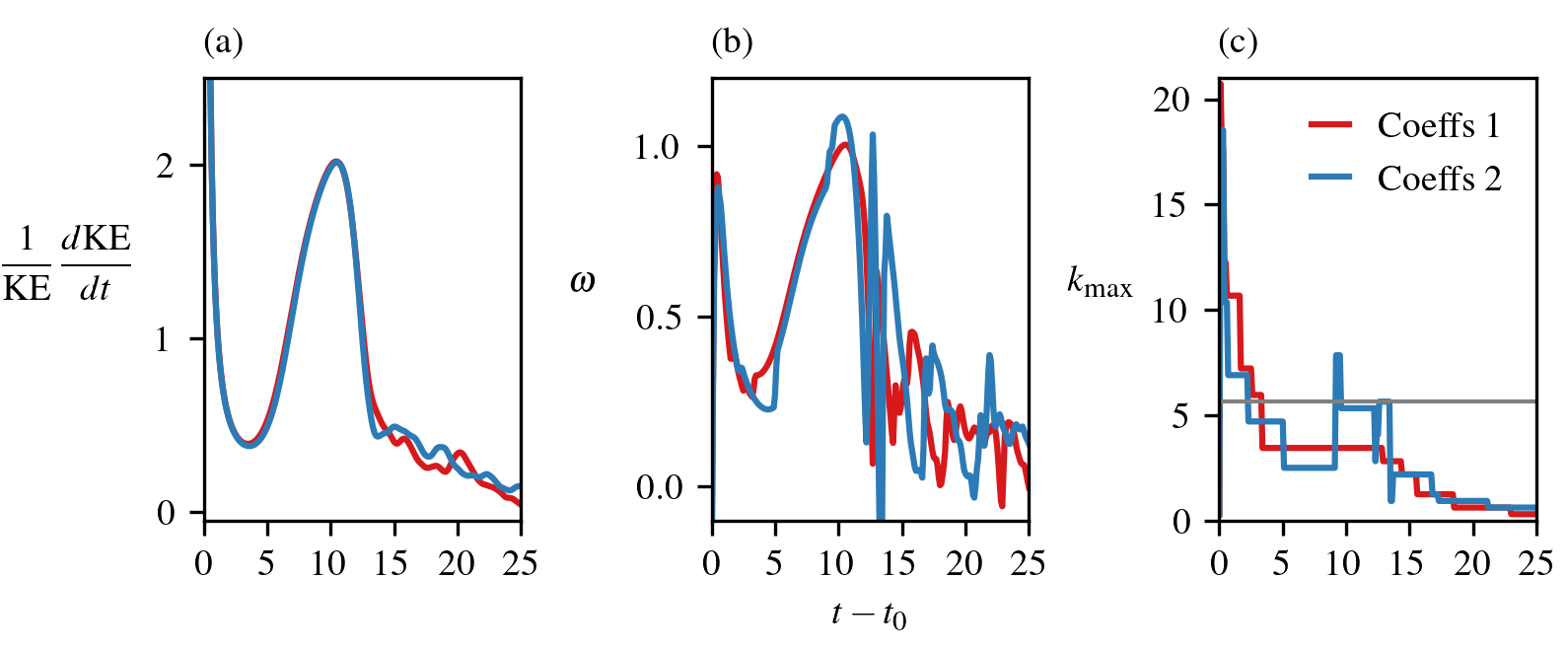}
    \caption{\label{fig:noise-coeffs} Sensitivity of results to the random coefficients ($c_{i,j}$) in the initial perturbations used in the numerical simulations with scale $\tilde{S}=10^{-2}$ for the cryomictic ($\lambda/H=1$, $T_H/\TMD=0.2$) regime and $\gamma=10^{-2}$, $\Ra=10^6$ and $\delta/H=0.2$. (a) The kinetic energy (KE, equation~\ref{eq:KEgrowth}) growth rate, (b) modal energy growth rate ($\omega$, equation~\ref{eq:omega}) and (c) most unstable wavenumber ($k_{\max}$, equation~\ref{eq:Gtk}) with the representative most unstable wavenumber, $\tilde{k}_{\max}(t_1)$, overlaid in grey.}
\end{figure}

In figures~\ref{fig:noise-scale} and \ref{fig:noise-coeffs}, the sensitivity of results to the initial scale ($\tilde{S}$, figure~\ref{fig:noise-scale}) and random coefficients ($c_{i,j}$, figure~\ref{fig:noise-coeffs}) of the initial perturbations \eqref{eq:initial_perts} used in the numerical simulations can be seen. As before, this is performed on the cryomictic case since RTIs begin to grow immediately. 

The time of RTI formation ($t_1$) is moderately sensitive to the scale ($\tilde{S}$) where decreases in $\tilde{S}$ leads to longer $t_1$ (figure~\ref{fig:noise-scale}a and b). Specifically, increasing from $t_1\approx9$ at $\tilde{S}=10^{-1}$ to $t_1\approx18$ at $\tilde{S}=10^{-6}$. Conversely, changes in the initial perturbation structure by varying the random coefficients ($c_{i,j}$) do not impact measurements of $t_1$ (figure~\ref{fig:noise-coeffs}a). Note that we denote the random coefficients used throughout this work as ``Coeffs 1''. The growth rate ($\omega$) is generally insensitive to changes in both $\tilde{S}$ and $c_{i,j}$ for times $t\leq t_1$ (figures~\ref{fig:noise-scale}b and \ref{fig:noise-coeffs}d). Additionally, for $t\leq t_1$, the most amplified wavenumber ($k_{\max}$) is generally insensitive to changes in scale (figures~\ref{fig:noise-scale}c). Although $k_{\max}$ is sensitive to the initial structure, suggesting that over many simulations of different initial perturbations (i.e. varying $\tilde{S}$ and $c_{i,j}$), a distribution of most amplified wavenumbers would emerge. 

Lastly, we note that the sensitivity of results on $t_1$ with $\tilde{S}$ would impact the threshold used to predict $t_1$ in the linear analysis (section~\ref{sec:linear}) since the threshold of 1000 is chosen a posteriori. Clearly, this depends on the initial noise scale (figure~\ref{fig:noise-scale}c) but would be difficult to find a priori. 
Nevertheless, the estimated $t_1$ is still reliable since the observed $t_1$ from the simulations is relatively insensitive to the initial perturbation amplitude. Further, the linear analysis still captures the early growth dynamics quite well, namely the amplification, growth rate and most unstable wavenumber. 

%%%%%%%%%%%%%%%%%%%%%%%%%%%%%%%%%%%%%%%%%%%%%%%%%%%%%%%%%%%%%%%%
%%%%%%%%%%%%%%%%%%%%%%%%%%%%%%%%%%%%%%%%%%%%%%%%%%%%%%%%%%%%%%%%
\section{Analytical Solution to the 1D Radiatively Forced Heat Equation}\label{app:analytic}
%%%%%%%%%%%%%%%%%%%%%%%%%%%%%%%%%%%%%%%%%%%%%%%%%%%%%%%%%%%%%%%%

Consider the one-dimensional form of the radiatively forced heat equation \eqref{eq:heatND} in the quiescent regime,
\begin{equation}\label{eq:1dheat_no-force}
    \frac{\partial T}{\partial t} = \frac{1}{\sqrt{\Ra\Pr}}\frac{\partial^2 T}{\partial z^2} + \frac{1}{\gamma\sqrt{\Ra\Pr}}\left(\frac{\lambda}{H}\right)^{1-\sigma}e^z.
\end{equation}
It is beneficial to rewrite this in terms of the forcing timescale via the transformation\\ $t \rightarrow (\tau/\tau_f)t$ (i.e. $t$ now refers to the forcing time $t=t^*/\tau_f$) so that
\begin{equation}\label{eq:1dtemp}
    \frac{\partial T}{\partial t} = \gamma\frac{\partial^2 T}{\partial z^2} + e^z.
\end{equation}
Without loss of generality, we have also rescaled the temperature by setting $(\lambda/H)^{1-\sigma}\equiv1$ and dropping the subtraction of $\TMD$ since (i) they can be restored later once the solution has been obtained, and (ii) it guarantees the condition $T(z=0)=0$, so that the initial state becomes 
\begin{equation}\label{eq:forcingIC}
    T(t=0) = T_0 = \frac{1}{2}\int_z^0 1 + \tanh{\left(\frac{c_1z'+1}{c_2}\right)}\; dz'.
\end{equation}
The boundary conditions are then given by
\begin{equation}
    T(z=0) = \lim_{z\rightarrow-\infty} T-T_0 = 0,
\end{equation}
where the latter refers to the far-field condition that far away from the ice-cover, the water remains undisturbed. 

Equation~\eqref{eq:1dtemp} was originally solved by \citet{MATTHEWS1987} for a linear stratification (i.e. $T_0=A+Bz$). Here, we adapt their analysis to construct a new solution for any given initial profile. Following, \citet{MATTHEWS1987}, consider the Laplace transform of equation \eqref{eq:1dtemp} which upon rearranging gives
\begin{equation}\label{eq:Lap1dtemp}
    \frac{\partial^2 \breve{T}}{\partial z^2} - \frac{s}{\gamma}\breve{T} = -\frac{1}{\gamma}\left(\frac{e^z}{s} + T_0\right) 
\end{equation}
where $s$ is the Laplace transform variable and $\breve{T}(s,z)=\mathcal{L}\{T(t,z)\}$ is the Laplace transform of $T$. The advantage of this is that the PDE is effectively transformed into an ODE since only $z$ derivatives remain. The boundary conditions transform to 
\begin{equation}\label{eq:LapBCs}
    \breve{T}(z=0)= \lim_{z\rightarrow-\infty} \breve{T} - \frac{T_0}{s} = 0.
\end{equation}
The general solution to \eqref{eq:Lap1dtemp} can be calculated using variation of parameters to give
\begin{equation}
    \breve{T}(s,z) = \left(C_1 + \int_{z}^0\frac{e^{-\sqrt{\frac{s}{\lambda}}\zeta}}{2\sqrt{s\gamma}}T_0(\zeta)\;d\zeta\right)e^{\sqrt{\frac{s}{\lambda}}z} + \left(C_2 - \int_{z}^0\frac{e^{\sqrt{\frac{s}{\lambda}}\zeta}}{2\sqrt{s\gamma}}T_0(\zeta)\;d\zeta\right)e^{-\sqrt{\frac{s}{\lambda}}z} + \frac{e^z}{s(s-\gamma)}
\end{equation}
where $C_1$ and $C_2$ are constants of integration that are determined by \eqref{eq:LapBCs},
\begin{equation}
    C_1 = -\frac{1}{s(s-\gamma)} - \int_{-\infty}^0 \frac{e^{\sqrt{\frac{s}{\lambda}}z}}{2\sqrt{s\gamma}}T_0(z)\;dz, \qquad C_2 = \int_{-\infty}^0 \frac{e^{\sqrt{\frac{s}{\lambda}}z}}{2\sqrt{s\gamma}}T_0(z)\;dz.
\end{equation}
The solution can be written as
\begin{equation}
    \breve{T}(s,z) = \frac{1}{2\sqrt{s\gamma}}\int_{-\infty}^0\left(e^{-\sqrt{\frac{s}{\lambda}}|z-\zeta|} - e^{-\sqrt{\frac{s}{\lambda}}|z+\zeta|}\right)T_0(\zeta)\;d\zeta + \frac{e^z-e^{\sqrt{\frac{s}{\gamma}}z}}{s(s-\gamma)},
\end{equation}
and by taking the inverse Laplace transform, 
\begin{align}
    T(t,z) = &\int_{-\infty}^0\left(\mathcal{L}^{-1}\left\{\frac{e^{-\sqrt{\frac{s}{\lambda}}|z-\zeta|}}{2\sqrt{s\gamma}}\right\} - \mathcal{L}^{-1}\left\{\frac{e^{-\sqrt{\frac{s}{\lambda}}|z+\zeta|}}{2\sqrt{s\gamma}}\right\}\right)T_0(\zeta)\;d\zeta \nonumber \\[5pt]
    &\hspace*{4cm} + e^z\mathcal{L}^{-1}\left\{\frac{1}{s(s-\gamma)}\right\} - \mathcal{L}^{-1}\left\{\frac{e^{-\sqrt{\frac{s}{\gamma}}|z|}}{s(s-\gamma)}\right\}.
\end{align}
The first three inverse transforms are tabulated \citep[e.g.][]{spiegel1992} to give
\begin{equation}
    \mathcal{L}^{-1}\left\{\frac{e^{-\sqrt{\frac{s}{\lambda}}|z\pm\zeta|}}{2\sqrt{s\gamma}}\right\} = \frac{e^{-\frac{(z\pm\zeta)^2}{4\gamma t}}}{2\sqrt{\pi\gamma t}}, \qquad \mathcal{L}^{-1}\left\{\frac{1}{s(s-\gamma)}\right\} = \frac{e^{\gamma t}-1}{\gamma},
\end{equation}
whilst the last can be calculated using the convolution theorem,
\begin{equation}
    \mathcal{L}^{-1}\left\{\frac{e^{-\sqrt{\frac{s}{\gamma}}|z|}}{s(s-\gamma)}\right\} = -\frac{e^{\gamma t}-1}{\gamma}*\frac{ze^{-\frac{z^2}{4\gamma t}}}{2\sqrt{\pi\gamma t^3}} = -\frac{z}{2\sqrt{\pi\gamma^3}}\int_0^t\frac{e^{\gamma (t-\xi)}-1}{\xi^\frac{3}{2}}e^{-\frac{z^2}{4\gamma\xi}}\; d\xi.
\end{equation}
After some manipulation, the temperature evolution in the quiescent regime is given by
\begin{align}\label{eq:1Dsol}
    T(t,z) = & \frac{1}{2\sqrt{\pi\gamma t}}\int_{-\infty}^0\left(e^{-\frac{(z-\zeta)^2}{4\gamma t}} - e^{-\frac{(z+\zeta)^2}{4\gamma t}}\right)T_0(\zeta)\; d\zeta + \frac{1}{\gamma}\erfc{-\frac{z}{2\sqrt{\gamma t}}} - \frac{e^z}{\gamma} \nonumber \\[5pt]
    &\hspace*{2cm} + \frac{e^{\gamma t}}{2\gamma}\left(e^z\erfc{\sqrt{\gamma t}+\frac{z}{2\sqrt{\gamma t}}} - e^{-z}\erfc{\sqrt{\gamma t}-\frac{z}{2\sqrt{\gamma t}}}\right),
\end{align}
where $\erfc{\cdot}$ is the complimentary error function. 

%%%%%%%%%%%%%%%%%%%%%%%%%%%%%%%%%%%%%%%%%%%%%%%%%%%%%%%%%%%%%%%%
%%%%%%%%%%%%%%%%%%%%%%%%%%%%%%%%%%%%%%%%%%%%%%%%%%%%%%%%%%%%%%%%
\section{Chebyshev Spectral Collocation \& Null Space Method}\label{app:LA-methods}
%%%%%%%%%%%%%%%%%%%%%%%%%%%%%%%%%%%%%%%%%%%%%%%%%%%%%%%%%%%%%%%%
\subsection{Chebyshev Spectral Collocation}\label{app:cheby}

To first apply the Chebyshev collocation method, the domain is truncated to the finite interval $z\in[-Z,0]$ where $Z$ is sufficiently large enough such that further increases do not significantly affect the solution. For similar reasons as for the simulations in section~\ref{sec:dns}, $Z=-10$. To apply the expansion of Chebyshev polynomials, the truncated domain is mapped onto the interval $\xi\in[-1,1]$ via the linear transformation
\begin{equation}
    z=\frac{Z}{2}(\xi-1),
\end{equation}
where the boundary conditions \eqref{eq:2D_perts_BCs} are imposed at the endpoints $\xi=\pm1$. The modal functions are approximated by a set of truncated Chebyshev series of the form
\begin{equation}\label{eq:Cheby_expans}
    \hat{q}(\xi,t) = \sum_{n=0}^{N} a_{n}^{(q)}(t)\, T_n(\xi).
\end{equation}
Here, $q$ represents the flow variables $u$, $w$, $P$ or $T$ (e.g. $a_n^{(w)}$ are the expansion coefficients for the vertical velocity mode $\hat{w}$); $T_n(\xi)$ denotes the Chebyshev polynomial of the first kind of degree $n$; and $N$ is the highest order polynomial used. 

The domain is discretised into $N+1$ Gauss-Lobatto collocation points
\begin{equation}
\xi_j=\cos\left(\frac{j\pi}{N}\right), \qquad j=0,1,\ldots,N,
\end{equation}
where at the endpoints ($j=0,N$) the boundary conditions \eqref{eq:2D_perts_BCs} are applied. Upon discretisation, the boundary conditions and perturbation equations \eqref{eq:2D_pert_eqns} can be written in matrix form as
\begin{equation}\label{eq:matrix_eqns}
    \mat{C}\vec{a}_s = 0, \qquad \mat{A}\frac{d\vec{a}_s}{dt}=\mat{B}(t)\vec{a}_s + \mat{G}\vec{a}_p,
\end{equation}
where the time-dependent coefficients are collected into the vectors
\begin{gather} \nonumber
    \vec{a}_{s}(t) =
    \left[
    a_{0}^{(u)}, \,\ldots\,, \,a_{N}^{(u)},\,
    a_{0}^{(w)}, \,\ldots\,, \,a_{N}^{(w)},\,
    a_{0}^{(T)}, \,\ldots\,, \,a_{N}^{(T)}
    \right]^\text{T}, \\
    \vec{a}_{P}(t)=
    [a_{0}^{(P)}, \,\ldots\,, \,a_{N}^{(P)}]^\text{T}.
\end{gather}
The matrix $\mat{C}$ (supplementary material S2.1) represents the boundary and incompressibility conditions whereas the matrices $\mat{A}$, $\mat{B}(t)$ and $\mat{G}$ (supplementary material S2.2) represent the momentum and heat equations enforced on the interior points ($j=1,2,...,N-1$).

\subsection{Null Space Method}\label{app:null}

Since the boundary conditions \eqref{eq:2D_perts_BCs} are homogeneous, then $\mat{C}\vec{a}_s\equiv0$, and $\vec{a}_s$ can be expressed in terms of vectors that span the null space of $\mat{C}$ (supplementary material S2.1). In other words,
\begin{equation}\label{eq:reduceNC}
    \vec{a}_{s}(t)=\mat{N}_\mat{C}\vec{b}_s(t),
\end{equation}
where the columns of $\mat{N}_\mat{C}$ forms an orthonormal basis of $\text{Null}(\mat{C})$ with coordinates $\vec{b}_s$. Equivalently, each column of $\mat{N}_{\mat{C}}$ defines, through the expansions in \eqref{eq:Cheby_expans}, a mode for $\hat{u}$, $\hat{w}$, and $\hat{T}$ that satisfies both the incompressibility and boundary conditions, and so the vector $\vec{b}_s$ contains the modal coordinates. For brevity, the construction of $\mat{N}_{\mat{C}}$ is given in supplementary material S2.1. Substituting equation \eqref{eq:reduceNC} into \eqref{eq:matrix_eqns} and upon rearranging yields the augmented system
\begin{equation}\label{eq:augcoll}
    \begin{bmatrix}
        \mat{A}\mat{N}_\mat{C} & -\mat{G}
    \end{bmatrix}
    \begin{bmatrix}
        d\vec{b}_s/dt \\
        \vec{a}_{p}(t)
    \end{bmatrix} =
    \mat{B}(t)\mat{N}_\mat{C}\vec{b}_{s}(t).
\end{equation}
Since the coefficient matrix on the left hand side is square and, typically, nonsingular, the system may be inverted to express $d\vec{b}_{s}/dt$ in terms of $\vec{b}_s$ as in equation~\eqref{eq:bstprime}. The matrix $\mat{Q}(t)$ is obtained from the inverse of the coefficient matrix and its construction is given in supplementary material S2.3. The pressure coefficients $\vec{a}_{p}(t)$ do not possess an independent evolution equation; rather, they act as Lagrange multipliers that enforce the incompressibility constraint at each instant.

%%%%%%%%%%%%%%%%%%%%%%%%%%%%%%%%%%%%%%%%%%%%%%%%%%%%%%%%%%%%%%%%
\bibliographystyle{jfm}
\bibliography{RDC.bib}
%%%%%%%%%%%%%%%%%%%%%%%%%%%%%%%%%%%%%%%%%%%%%%%%%%%%%%%%%%%%%%%%
%%%%%%%%%%%%%%%%%%%%%%%%%%%%%%%%%%%%%%%%%%%%%%%%%%%%%%%%%%%%%%%%

\end{document}

%%%%%%%%%%%%%%%%%%%%%%%%%%%%%%%%%%%%%%%%%%%%%%%%%%%%%%%%%%%%%%%%

\maketitle
\fontsize{12pt}{16pt}\selectfont
\setlength{\parindent}{0pt}

\newpage
%%%%%%%%%%%%%%%%%%%%%%%%%%%%%%%%%%%%%%%%%%%%%%%%%%%%%%%%%%%%%%%%
%%%%%%%%%%%%%%%%%%%%%%%%%%%%%%%%%%%%%%%%%%%%%%%%%%%%%%%%%%%%%%%%
\section{Reduction of the 3D Perturbation Equations to 2D}
%%%%%%%%%%%%%%%%%%%%%%%%%%%%%%%%%%%%%%%%%%%%%%%%%%%%%%%%%%%%%%%%

Consider a three-dimensional fourier mode decomposition of the perturbations:
\begin{equation}\label{eq:3D_fourier}
[\tilde{u}, \tilde{v}, \tilde{w}, \tilde{P}, \tilde{T}]^\text{T} = [\hat{u}(z,t), \hat{v}(z,t), \hat{w}(z,t), \hat{P}(z,t), \hat{T}(z,t)]^\text{T}e^{i(k_x x + k_y y)},
\end{equation}
where $v$ denotes the $y$-component of the velocity, and $k_x$ and $k_y$ are the wavenumbers in the $x$ and $y$ directions respectively. Further, consider a rotated coordinate system $(\chi_\parallel,\chi_\perp)$ in the $x-y$ plane where $\chi_\parallel$ and $\chi_\perp$ are components that are parallel and perpendicular, respectively, to the wavevector $(k_x,k_y)$, and given by
\begin{equation}\label{eq:rot_coord}
\chi_\parallel=\frac{k_x x + k_y y}{\sqrt{k_x^2+k_y^2}},
\qquad
\chi_\perp =\frac{k_x y - k_y x}{\sqrt{k_x^2+k_y^2}}.
\end{equation}
After transforming, the Fourier decomposition becomes
\begin{equation}\label{eq:fourier_rot}
[\upsilon_\parallel, \upsilon_\perp, w, P, T]^\text{T} =
[\hat{\upsilon}_\parallel(z,t), \hat{\upsilon}_\perp(z,t),\hat{w}(z,t), \hat{P}(z,t), \hat{T}(z,t)]e^{i\sqrt{k_x^2+k_y^2}\,\chi_\parallel},
\end{equation}
where $\upsilon_\parallel$ and $\upsilon_\perp$ denote the velocity components in the parallel ($\chi_\parallel$) and perpendicular ($\chi_\perp$) directions, respectively. Inserting these into the linearised equations (5.4)
% \eqref{eq:2D_pert_eqns}
gives
\begin{subequations}\label{eq:rotated_pert_eqns}
    \begin{gather}
        i\sqrt{k_x^2+k_y^2}\,\hat{\upsilon}_\parallel + \frac{\partial \hat{w}}{\partial z}=0, \\[5pt]
        \frac{\partial \hat{\upsilon}_\parallel}{\partial t} = -i\sqrt{k_x^2+k_y^2}\,\hat{P} + \sqrt{\frac{\Pr}{\Ra}}\left(\frac{\partial^2\hat{\upsilon}_\parallel}{\partial z^2} - (k_x^2+k_y^2)\hat{\upsilon}_\parallel\right), \\[5pt]
        \frac{\partial \hat{\upsilon}_\perp}{\partial t} = \sqrt{\frac{\Pr}{\Ra}}\left(\frac{\partial^2\hat{\upsilon}_\perp}{\partial z^2} - (k_x^2+k_y^2)\hat{\upsilon}_\perp\right), \\[5pt]
        \frac{\partial \hat{w}}{\partial t} = -\frac{\partial \hat{P}}{\partial z} + \sqrt{\frac{\Pr}{\Ra}}\left(\frac{\partial^2\hat{w}}{\partial z^2} - (k_x^2+k_y^2)\hat{w} \right) - \beta\bar{T}\hat{T}, \\[5pt]
        \frac{\partial \hat{T}}{\partial t} + \left(\frac{\lambda}{H}\right)^{\sigma-1}\gamma\sqrt{\Ra\Pr}\,\hat{w}\,\frac{\partial\bar{T}}{\partial z} = \frac{1}{\sqrt{\Ra\Pr}}\left(\frac{\partial^2\hat{T}}{\partial z^{*2}} - (k_x^2+k_y^2)\hat{T}\right).
    \end{gather}
\end{subequations}
It can be seen immediately that the perpendicular velocity $\hat{\upsilon}_\perp$ is decoupled and does not contribute to RTI development whereas the remaining coupled equations form a closed system that is identical in structure to the two-dimensional formulation (5.4). 
%\eqref{eq:2D_pert_eqns}
In particular, the horizontal wavenumber $k_y$ only appears in conjunction with $\sqrt{k_x^2+k_y^2}$, so that each three-dimensional mode with wavevector $(k_x,k_y)$ is dynamically equivalent to a two-dimensional mode with wavenumber $\sqrt{k_x^2+k_y^2}$. Therefore, the three-dimensional problem reduces to an equivalent two-dimensional problem. Note that this holds only for horizontally periodic domains with no horizontal variations in the base state.

%%%%%%%%%%%%%%%%%%%%%%%%%%%%%%%%%%%%%%%%%%%%%%%%%%%%%%%%%%%%%%%%
%%%%%%%%%%%%%%%%%%%%%%%%%%%%%%%%%%%%%%%%%%%%%%%%%%%%%%%%%%%%%%%%
\newpage
\section{Construction of the Chebyshev Collocation Matrices}
%%%%%%%%%%%%%%%%%%%%%%%%%%%%%%%%%%%%%%%%%%%%%%%%%%%%%%%%%%%%%%%%
Here, the Chebyshev evaluation matrices used to impose the
constraints and to discretise the linearised equations (5.4) are defined.
%\eqref{eq:2D_pert_eqns}

\subsection{Construction of the Null Space Matrix ($\mat N_{\mat C}$)}

At the interior Gauss-Lobatto collocation points ($\xi_j:j=1,\ldots,N-1$), let
\begin{equation}
    (\mat T)_{j,n}=T_n(\xi_j), \qquad
    (\mat T')_{j,n}
    =
    \left.\frac{dT_n}{d\xi}\right|_{\xi=\xi_j},
    \qquad
    (\mat T'')_{j,n}
    =
    \left.\frac{d^2T_n}{d\xi^2}\right|_{\xi=\xi_j},
\end{equation}
for $n=0,\ldots,N$, where $T_n$ are Chebyshev polynomials of the first kind. At the boundary points $\xi_0=1$ and $\xi_N=-1$, let
\begin{equation}
    \mat T_0 = T_n(\xi_0),
    \quad
    \mat T'_0 =
    \left.\frac{dT_n}{d\xi}\right|_{\xi=\xi_0},
    \quad
    \mat T_N = T_n(\xi_N),
    \quad
    \mat T'_N =
    \left.\frac{dT_n}{d\xi}\right|_{\xi=\xi_N}.
\end{equation}

Using the mapping $z=Z(\xi-1)/2$, the $z$ derivatives can be written as
\begin{equation}
    \mat D_z = \frac{2}{L}\mat T',
    \qquad
    \mat D_{zz} =
    \left(\frac{2}{L}\right)^2\mat T''.
\end{equation}

By defining $\vec{a}_s = [\vec{a}^{(u)}, \vec{a}^{(w)}, \vec{a}^{(T)}]^\text{T}$, the boundary and incompressibility constraints are written as 
\begin{equation}\label{SMeq:homogC}
    \mat C\vec a_s=0
\end{equation}
where $\mat C\in\mathbb{C}^{(N+7)\times (3N+3)}$ is given by
\begin{equation}
    \mat{C} = 
    \begin{bmatrix}
        \mat T_N & \mat 0 & \mat 0 \\
        \mat 0 & \mat T_N & \mat 0 \\
        \mat 0 & \mat 0 & \mat T_N \\
        \mat 0 & \mat T'_N & \mat 0 \\
        \mat T_0 & \mat 0 & \mat 0 \\
        \mat 0 & \mat T_0 & \mat 0 \\
        \mat 0 & \mat 0 & \mat T_0 \\
        \mat 0 & \mat T'_0 & \mat 0 \\
        ik\mat T & \mat D_z & \mat 0
    \end{bmatrix}.
\end{equation}
The top eight block rows impose the homogeneous boundary conditions (5.5) 
%\eqref{eq:2D_perts_BCs}
whilst the final block row imposes the incompressibility constraint (5.4a)
%\eqref{eq:2D_pert_eqns_cont}
on the interior collocation points. \\

Since $\vec{a}_s$ must satisfy equation~\eqref{SMeq:homogC}, then $\vec{a}_s$ must lie in the null space of $\mat C$, i.e. the subspace spanned by the orthonormal basis $\mat N_{\mat C}\in\mathbb{C}^{(3N+3)\times(2N-4)}$. Hence, for any state satisfying the discrete boundary and incompressibility constraints \eqref{SMeq:homogC}, 
\begin{equation}
    \vec a_s(t)=\mat N_{\mat C}\vec b_s(t),
\end{equation}
where $\vec{b}_s$ represents the coordinates within the null space or, equivalently, the set of independent modal coefficients. In practice, $\mat N_{\mat C}$ is computed using the singular value
decomposition of $\mat C=\mat U \mat S \mat V^{\mathrm{T}}$ and the columns of $\mat V$ corresponding to zero singular values are collected. 

%%%%%%%%%%%%%%%%%%%%%%%%%%%%%%%%%%%%%%%%%%%%%%%%%%%%%%%%%%%%%%%%
\newpage
\subsection{Construction of $\mat{A}$, $\mat{B}(t)$, and $\mat{G}$}

The matrices $\mat A$, $\mat B(t)$, and $\mat G$ are obtained by enforcing the momentum and heat equations (5.4b-d)
%\eqref{eq:2D_pert_eqns}
at the interior collocation points. To account for the background temperature field, the square diagonal matrices of size $N-1$ are defined,
\begin{equation}
    \mat{U}(t) = \mathrm{diag}
    \left(\bar{T}(\xi_1,t),\, \ldots\, , \,\bar{T}(\xi_{N-1},t) \right), 
    \;
    \mat{U}_z(t) = \mathrm{diag}
    \left(
    \left.\frac{\partial\bar{T}}{\partial z}\right|_{\xi=\xi_1},\, \ldots\,, \,
    \left.\frac{\partial\bar{T}}{\partial z}\right|_{\xi=\xi_{N-1}}
    \right).
\end{equation}
This allows the system to be expressed in matrix form as
\begin{equation}
    \mat{A} =
    \begin{bmatrix}
        \mat{T} & \mat{0} & \mat{0}\\
        \mat{0} & \mat{T} & \mat{0}\\
        \mat{0} & \mat{0} & \mat{T}
    \end{bmatrix},
\end{equation}
\begin{equation}
    \mat{B}(t)=
    \begin{bmatrix}
        \sqrt{\Pr/\Ra}\,\bigl(\mat{D}_{zz}-k^2\mat{T}\bigr) & \mat{0} & \mat{0} \\[3pt]
        \mat{0} & \sqrt{\Pr/\Ra}\,\bigl(\mat{D}_{zz}-k^2\mat{T}\bigr) & \beta\mat{U}(t)\mat{T} \\[3pt]
        \mat{0} & -(\lambda/H)^{\sigma-1}\gamma\sqrt{\Ra\Pr}\mat{U}_z(t)\mat{T} & \bigl(\mat{D}_{zz}-k^2\mat{T}\bigr)/\sqrt{\Ra\Pr}
    \end{bmatrix},
\end{equation}
and
\begin{equation}
    \mat{G}=
    \begin{bmatrix}
        -ik\mat{T} \\
        -\mat{D}_z \\
        \mat{0}
    \end{bmatrix},
\end{equation}
where $\mat{0}$ is the zero matrix of size $N-1\times N+1$.

%%%%%%%%%%%%%%%%%%%%%%%%%%%%%%%%%%%%%%%%%%%%%%%%%%%%%%%%%%%%%%%%

\subsection{Construction of $\mat{Q}(t)$}

By introducing the matrix
\begin{equation}
    \mat{K} =
    \begin{bmatrix}
        \mat{A}\mat{N}_\mat{C} & -\mat{G}
    \end{bmatrix},
\end{equation}
the solution to (5.6) can be written as 
%\eqref{eq:bstprime}
as
\begin{equation}
    \begin{bmatrix}
    d\vec{b}_s/dt\\
    \vec{a}_p
    \end{bmatrix}
    = \mat{K}^{-1}\mat{B}(t)\mat{N}_\mat{C}\vec{b}_s(t)
\end{equation}
where $\mat{K}^{-1}$ is the left inverse of $\mat{K}$. Extracting the first block yields
\begin{equation}
    \frac{d\vec{b}_s}{dt} = \mat{Q}(t)\,\vec{b}_s,
\end{equation}
where $\mat{Q}(t)$ is given by
\begin{equation}
    \mat{Q}(t) = \mat{P}\mat{K}^{-1}\mat{B}(t)\mat{N}_\mat{C},
\end{equation}
and $\mat{P}$ is a constant matrix that extracts the components associated with $d\vec{b}_s/dt$. \\

It should be noted that $\mat{K}$ depends only on the discretisation and the wavenumber $k$, and is therefore time-independent. Thus, the inverse is only computed once, since the time dependence of $\mat{Q}$ arises solely from $\mat{B}(t)$, and avoids repeated matrix inversion.

%%%%%%%%%%%%%%%%%%%%%%%%%%%%%%%%%%%%%%%%%%%%%%%%%%%%%%%%%%%%%%%%
%%%%%%%%%%%%%%%%%%%%%%%%%%%%%%%%%%%%%%%%%%%%%%%%%%%%%%%%%%%%%%%%